%% file: main.tex
\documentclass[runningheads]{llncs}

\usepackage[T1]{fontenc}
\usepackage{graphicx}

\usepackage{amsmath}
\usepackage{amsfonts}
\usepackage{stmaryrd}
\usepackage{bussproofs}
\usepackage{stackengine}
\usepackage[most]{tcolorbox}
\usepackage{thm-restate}
\usepackage[ruled,vlined,linesnumbered]{algorithm2e}
\usepackage{cleveref}

\usepackage{tikz}
\usepackage{pgfplots}
\usepackage{subcaption}
\usetikzlibrary{arrows,automata,positioning,calc}

\usepackage{bbding}

\usepackage{subcaption}
\input{macros}

\begin{document}

\title{Modular Attractor Acceleration\\ in Infinite-State Games (Full Version)}
\author{
    Philippe Heim\Envelope\orcidID{0000-0002-5433-8133} \and
    Rayna Dimitrova\orcidID{0009-0006-2494-8690}
}
\authorrunning{P. Heim \and R. Dimitrova}
\institute{
    CISPA Helmholtz Center for Information Security, Saarbr\"ucken, Germany
    \email{\{philippe.heim, dimitrova\}@cispa.de}
}

\maketitle

\begin{abstract}
Infinite-state games provide a framework for the synthesis of reactive systems with unbounded data domains. Solving such games typically relies on computing symbolic fixpoints, particularly symbolic attractors. However, these computations may not terminate, and while recent acceleration techniques have been proposed to address this issue, they often rely on acceleration arguments of limited expressiveness.

In this work, we propose an approach for the modular computation of acceleration arguments. It enables the construction of complex acceleration arguments by composing simpler ones, thereby improving both scalability and flexibility. In addition, we introduce a summarization technique that generalizes discovered acceleration arguments, allowing them to be efficiently reused across multiple contexts. Together, these contributions improve the efficiency of solving infinite-state games in reactive synthesis, as demonstrated by our experimental evaluation.

\end{abstract}

\section{Introduction}\label{sec:intro}
\input{intro}

\section{Preliminaries}\label{sec:games}
\input{games}

\section{Generalized Acceleration Lemmas}\label{sec:lemmas}
\input{lemmas}

\section{Enforcement Summaries}\label{sec:summaries}
\input{summaries}

\section{Implementation and Experimental Evaluation}\label{sec:experiments}
\input{experiments}

\begin{credits}

\subsubsection{Data Availability Statement.}
The software generated during and analysed during the current study, as well as the associated data and benchmarks, is available in the Zenodo repository~\url{https://doi.org/10.5281/zenodo.18163658}.

\subsubsection{\discintname}
The authors have no competing interests to declare that are relevant to the content of this article.
\end{credits}

\bibliographystyle{splncs04}
\bibliography{%
    bib/logic-sat-automata.bib,%
    bib/own-publications.bib,%
    bib/smt-chc-sygus-co.bib,%
    bib/synthesis-abstraction.bib,%
    bib/synthesis-finite.bib,%
    bib/synthesis-other.bib,%
    bib/synthesis-symbolic.bib,%
    bib/synthesis-theory.bib,%
    bib/textbooks.bib,%
    bib/verification-control-monitoring.bib,%
    bib/verification-termination.bib%
}

\appendix

\newpage

\section{Proofs}\label{app:proofs}
\input{appendix-proofs}

\section{Additional Data}\label{app:additional-data}
\input{appendix-more-data}

\end{document}

%% file: macros.tex
\newcommand{\issy}{\texttt{Issy}}
\newcommand{\sweap}{\texttt{sweap}}
\newcommand{\syntheos}{\texttt{Syntheos}}
\newcommand{\muval}{\texttt{MuVal}}

\newcommand{\be}[1]{\textbf{#1}}

\newcommand{\Nat}{\ensuremath{\mathbb{N}}}

\newcommand{\Real}{\ensuremath{\mathbb{R}}}

\newcommand{\sema}[1]{\llbracket #1 \rrbracket}

\newcommand{\values}{\ensuremath{\mathcal{V}}}

\newcommand{\vars}{\ensuremath{\mathit{Vars}}}

\newcommand{\assignment}{\ensuremath{\nu}}
\newcommand{\assignments}[1]{\ensuremath{\mathit{Assignments}}(#1)}

\newcommand{\assmt}[1]{\mathbf{#1}}

\newcommand{\FOL}[1]{\ensuremath{\mathit{FOL}}(#1)}
\newcommand{\FOLX}{\ensuremath{\mathit{FOL}}}
\newcommand{\QF}[1]{\ensuremath{\mathit{QF}}(#1)}
\newcommand{\QFX}{\ensuremath{\mathit{QF}}}
\newcommand{\FOLentails}{\ensuremath{\models}}
\newcommand{\FOLentailsT}[1]{\ensuremath{\models_{#1}}}

\newcommand{\states}{\ensuremath{\mathcal{S}}}
\newcommand{\symstates}{\ensuremath{\mathcal{D}}}

\newcommand{\cpre}[2]{\ensuremath{\mathit{CPre}_{#1}(#2)}}
\newcommand{\cpreX}[1]{\ensuremath{\mathit{CPre}_{#1}}}

\newcommand{\sys}{\ensuremath{\mathit{Sys}}}
\newcommand{\env}{\ensuremath{\mathit{Env}}}

\newcommand{\str}[2]{\mathit{Strat}_{#1}(#2)}

\newcommand{\plays}{\mathit{Plays}}

\newcommand{\symgame}{\mathcal{G}}

\newcommand{\dom}{\ensuremath{\mathit{dom}}}

\newcommand{\linit}{l_{\mathit{init}}}

\newcommand{\validinput}{\mathit{ValidIn}}

\newcommand{\changed}[1]{\textcolor{blue}{#1}}

\newcommand{\progvars}{\ensuremath{\mathbb{X}}}
\newcommand{\inputs}{\ensuremath{\mathbb{I}}}

\newcommand{\inv}{\ensuremath{\mathit{inv}}}
\newcommand{\base}{\ensuremath{\mathit{base}}}
\newcommand{\conc}{\ensuremath{\mathit{conc}}}
\newcommand{\step}{\ensuremath{\mathit{step}}}
\newcommand{\stay}{\ensuremath{\mathit{stay}}}

\newcommand{\loopgame}{\ensuremath{\mathsf{LoopGame}}}

\newcommand{\locsplit}{\ensuremath{l_\mathit{Split}}}
\newcommand{\locend}{\ensuremath{l_\mathit{End}}}
\newcommand{\combine}[2]{\langle #1, #2 \rangle}

\newcommand{\nextp}{\mathit{next}}
\newcommand{\Summaries}{\mathit{Summaries}}

\newcommand{\metavars}{\mathit{Meta}}


%% file: intro.tex
Two-player infinite-duration graph games are a standard formalism used for synthesis of correct-by-construction reactive systems.
Synthesizing such a system reduces to computing a winning strategy for the system player against an adversarial environment.
Finite-state games are well-established, and efficient techniques and tools are available.
However, many systems operate over unbounded data domains, motivating interest in infinite-state games.
Solving such games is undecidable, and recent works explored several incomplete methods.
Abstraction-based methods~\cite{ChoiFPS22,MaderbacherB22,RodriguezS23,RodriguezS24,RodriguezGS24,AzzopardiSPS25,RodriguezGS25} reduce the problem to the finite case.
Their effectiveness depends on the abstraction domain's precision and the ability to discover relevant properties when refining the abstraction.
Constraint-based methods~\cite{FarzanK18,FaellaP23,SamuelDK23,MaderbacherWB24,HeimD24} operate directly on symbolic representations of the infinite state space.
The main challenge for these is taming the divergence of the computations.

As many game-solving algorithms rely on (nested) fixpoint computations, one line of recent work~\cite{HeimD24,SchmuckHDN24,HeimD25b} has focused on \emph{accelerating} these computations, especially \emph{accelerating attractor computations} which are the main driver of those fixpoint-based algorithms.
Intuitively, an attractor is the set of states in the game from which a given player has a strategy to reach a given target set of states, regardless of the opponent's behavior.
A key notion in~\cite{HeimD24,SchmuckHDN24,HeimD25b} is that of \emph{acceleration lemmas}, which help to establish the existence of strategies for infinite subsets of the attractor, by capturing the effect of an infinite number of computation steps into a single one.
Finding acceleration lemmas is challenging, as it requires proving that an acceleration property can be enforced against any adversarial behavior.
\cite{HeimD24} computes acceleration lemmas by representing candidates for them as uninterpreted functions and by using symbolic game solving to generate constraints on those.
However, the resulting constraint systems can become prohibitively large in complex games.
The burden of finding acceleration lemmas is then shifted entirely to the SMT solver.
Subsequent methods, such as~\cite{SchmuckHDN24} have improved this computation but have not departed conceptually from the original idea.
Overall, despite promising results, existing methods often struggle when complex acceleration lemmas are required.

In this work, we investigate methods for efficiently constructing complex acceleration lemmas by systematically composing them from simpler ones, which allows us to conduct a dedicated search for lemmas.
In addition, we explore how acceleration arguments can be reused through appropriate generalization.
The following examples motivate these ideas.

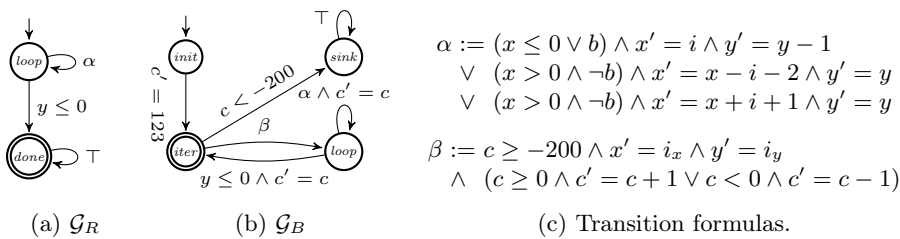
\begin{figure}[t!]
    \begin{subfigure}[b]{0.13\textwidth}
        \centering
        \input{example-reach}
        \vspace{2mm}
        \caption{$\mathcal G_R$}\label{fig:example-reach}
    \end{subfigure}
    \begin{subfigure}[b]{0.30\textwidth}
            \centering
        \input{example-buchi}
         \caption{$\mathcal G_B$}\label{fig:example-buchi}
    \end{subfigure}
      \begin{subfigure}[b]{0.54\textwidth}
            \centering
	$
	\begin{array}{lll} \small
	\alpha &  := &
	(x \leq 0 \lor b)\land x ' = i \land y' = y-1 \\
	& \lor & (x > 0\land \neg b) \land x ' = x - i - 2 \land y' = y \\
	& \lor & (x > 0\land \neg b) \land x ' = x + i +1  \land y' = y
 	\end{array}
	$

	\medskip
	
	$
	\begin{array}{lll} \small
	\beta &  := &
	c \geq -200 \land  x ' = i_x  \land y' = i_y \\
	& \land & (c \geq 0 \land c' = c+ 1 \lor c < 0 \land c' = c - 1)
 	\end{array}
	$

	\caption{Transition formulas.}\label{fig:trans-formula}
    \end{subfigure}
\caption{Symbolic games $\mathcal G_R$ and $\mathcal G_B$ for the motivating examples.  In both games,  $x$ and $y$ are integer state variables,  and $i$ and $b$ are input variables with integer and Boolean types, respectively. Additionally, the B\"uchi game $\mathcal G_B$ has an integer state variable $c$ and input variables $i_x$ and $i_y$, which are also integers. }\label{fig:intro-examle}
\end{figure}

\begin{example}\label{ex:intro-reach}
\Cref{fig:example-reach} depicts a game $\mathcal G_R$, in which the system player must ensure reaching  location $\mathit{done}$ against all possible behaviors of the environment. The game starts in location $\mathit{loop}$ with arbitrary initial values of the state variables $x$ and $y$. The system controls $x$ and $y$, subject to the transition formulas labeling the game’s edges. In each step, the environment first chooses values for the input variables $i$ and $b$, after which the system selects a successor location and state consistent with the corresponding transition formula.
For example, from location $\mathit{loop}$, the system may remain in $\mathit{loop}$ and choose $x$ and $y$ satisfying the transition formula $\alpha$ shown in \Cref{fig:trans-formula}. Concretely, it may set $x$ to $i$ and decrement $y$ when $x \leq 0 \vee b$ holds, or otherwise update $x$ in one of two ways while keeping $y$ unchanged.
To show that the system wins, one must establish a strategy that, for any initial values of $x$ and $y$ and any sequence of inputs $i$ and $b$ chosen by the environment, eventually reaches  $\mathit{done}$. This amounts to enforcing a state $(\mathit{loop}, v_x, v_y)$ with $v_y \le 0$.
Such a strategy exists:
When $x > 0 \land \neg b$ holds, the system can update $x$ depending on $i$ to ensure that $x$ decreases. Hence, while $x > 0$ and the environment keeps $b$ false, the system can reduce $x$, eventually enabling the disjunct that decrements $y$.
When $y$ is decremented, the environment picks a new value $i$ for $x$, and the process repeats. Because $y$ never increases, the first disjunct can only be applied finitely many times with $y > 0$, i.e., the system can enforce progress toward $\mathit{done}$.
Establishing this formally requires finding a correct strategy (i.e., the system program) and the corresponding acceleration argument.
\end{example}

The reasoning above relies on establishing a lexicographic acceleration argument together with a strategy for the system player that enforces it.
The argument is that either $y$ decreases, in which case it does not matter how the value of $x$ changes,  or the value of $y$ remains unchanged and the value of $x$ decreases. The latter case relies on the system correctly selecting the update to $x$ based on the current input $i$ provided by the environment.

Although this argument is conceptually simple, the automatic synthesis of such acceleration lemmas is challenging for existing methods.
Current tools for solving infinite-state games fail on this example due to the vast search space of potential acceleration arguments and strategies.
In this paper, we address this challenge by introducing a modular approach for constructing such a witness.
The key idea is to define \emph{composition operations that build up more complex acceleration lemmas from simpler ones}.
This enables a structured navigation of the search space, leading to the discovery of powerful acceleration lemmas beyond the reach of existing techniques.
However, scalability remains a challenge, particularly when solving complex games that require generating multiple acceleration lemmas. The following example illustrates this issue.

\begin{example}\label{ex:intro-buchi}
\Cref{fig:example-buchi} depicts a game $\mathcal G_B$ with a B\"uchi winning condition requiring that location $\mathit{iter}$ be visited infinitely often.
The game contains a sub-game (locations $\mathit{iter}$ and $\mathit{loop}$) structurally similar to $\mathcal G_R$ from \Cref{ex:intro-reach}, extended with a state variable $c$ that remains unchanged by the transitions originating in $\mathit{loop}$.
As argued in \Cref{ex:intro-reach}, location $\mathit{iter}$ can always be reached from $\mathit{loop}$ for any initial values of $x$ and $y$.
To ensure that the loop between $\mathit{iter}$ and $\mathit{loop}$ executes infinitely often, the state variable $c$ must remain at least $-200$. Inspecting the transition formula $\beta$ shown in \Cref{fig:trans-formula}, we see that starting from any non-negative value of $c$, this is indeed possible. Therefore, the system player has a winning strategy, since the transition from the initial location $\mathit{init}$ to $\mathit{iter}$ initializes $c$ to $123$.
 However, applying the symbolic method for solving infinite-state B\"uchi games from~\cite{HeimD24} requires establishing acceleration arguments for multiple state sets:  $c \ge -200$, $c \ge -199$, $c \ge -198$, and so on -- until reaching the fixpoint $c \ge 0$.
Although the reasoning in each case is identical, this repetition introduces substantial computational overhead.
To address this inefficiency, we introduce the notion of \emph{enforcement summaries}, which are witnesses ``parameterized''  by the target set that a player can enforce reaching.
Once computed, an enforcement summary can be applied multiple times during the game's solving by checking lightweight applicability conditions. This is especially helpful when solving B\"uchi games like $\mathcal G_B$,  where for each iteration of the outer fixpoint computation, we need to apply acceleration to enforce convergence of the inner fixpoint computation.
Using enforcement summaries, each of the remaining 200 inner fixpoints collapses into a single computation step, thereby improving efficiency.
\end{example}
To compute these parameterized summaries, our method uses generic templates with  parameters that describe a general set of multiple possible target sets that the player can enforce.
Applying the summaries reduces to instantiating those parameters.
The key is to ensure that the templates are sufficiently general.

\paragraph{Related work.}

Recent years have seen significant progress in the synthesis of infinite-state reactive systems, particularly in solving infinite-state games. For comprehensive overviews, we refer the reader to~\cite{HeimD24,HeimD25b}.
In this paper, we introduce a novel framework and techniques for attractor acceleration, building upon the symbolic methods developed in~\cite{HeimD24}. These methods have proven effective for solving games that require reasoning about unbounded loops, where a player must make strategic decisions in response to the opponent.
To improve scalability, \cite{SchmuckHDN24} employs strategy templates in finite abstractions to identify smaller sub-games, which helps solve the overall game.
This approach hinges on balancing abstraction size and the usefulness of the sub-games' results.
The tool in~\cite{HeimD25b} builds on the methods and ideas in ~\cite{HeimD24,SchmuckHDN24} and implements further techniques for acceleration lemma computation.
In contrast to abstraction-based approaches~\cite{ChoiFPS22,MaderbacherB22,RodriguezS23,RodriguezS24,RodriguezGS24,AzzopardiSPS25,RodriguezGS25}, which treat finite-state synthesis as a black box, symbolic methods such as those in~\cite{HeimD24,HeimD25b} enable tight integration of game solving and constraint solving, allowing reasoning about data directly within the game-solving procedure.

The synthesis of ranking functions~\cite{ColonS01,BagnaraMPZ12,LeikeH15,UrbanGK16,BorrallerasBLOR17,FedyukovichZG18,ZhuK24} has been extensively studied in software verification and termination analysis, and loop acceleration techniques~\cite{BardinFLP03,BardinFLS05,KroeningSTTW13,KincaidBCR19,FrohnG19,Frohn20} have been successfully applied to prove termination or non-termination of numerical programs.
Synthesizing acceleration lemmas for system synthesis differs fundamentally from these verification settings.
While verification techniques reason about a given program, synthesis requires existential quantification over possible implementations. In other words, verification witnesses describe properties of all behaviors of a \emph{given program} (possibly via overapproximation), whereas synthesis witnesses must establish the \emph{existence of a program} that enforces a desired property.
Summaries have also been employed to improve verification efficiency~\cite{SeryFS11,SolankiCLR24,PimpalkhareK24}.
Our enforcement summaries, however, serve a distinct purpose: rather than summarizing the behavior of a program, they capture a \emph{set of properties that can be enforced} by a player in the game, corresponding to families of strategies for that player.
A majority of the methods and tools in verification generate ranking functions within a specific class, for example, (combinations of) linear ranking functions,  possibly utilizing general templates or grammars~\cite{ColonS01,LeikeH15,FedyukovichZG18}. Similarly,  our methods generate acceleration lemmas constructed from building blocks that are linear inequalities, and we use linear templates as the base for our enforcement summaries.  However, our methods for computing them are based on symbolic game solving.

\paragraph{Contributions.}
We generalize \emph{acceleration lemmas}~\cite{HeimD24} to enable their modular construction from simple lemmas via \emph{composition operations} such as intersection, lexicographic combination, and chaining.
We present a method for generating those integrated into symbolic game-solving procedures.
In contrast to previous work, this method includes a dedicated lemma search procedure that enables more targeted generation of acceleration lemmas.
We further extend these procedures with the \emph{computation and application of enforcement summaries} to prevent the repeated computation of similar acceleration lemmas.
We implemented our methods on top of the open-source tool \issy~\cite{HeimD25b}.
Our extensive experimental evaluation demonstrates that the proposed techniques can solve games beyond the reach of current tools while remaining competitive with the state of the art.\looseness=-1

%% file: example-reach.tex
\begin{tikzpicture}[->,>=stealth',shorten >=1pt,auto,node distance=2.5cm, initial text=]\scriptsize
  \tikzstyle{every state}=[fill=none,draw=black,text=black,inner sep=1.5pt, minimum size=12pt,thick,scale=0.7]

   \node[state, initial above] (l)     {$\mathit{loop}$};
   \node[state, accepting] (d) [below of=l,yshift=.7cm] {$\mathit{done}$};

  \path (l) edge[loop right]  node {$\small \alpha$} (l);
  \path (l) edge  node[right] {$\small y \leq 0$} (d);
  \path (d) edge[loop right]  node {$\small \top$} (d);
\end{tikzpicture}

%% file: example-buchi.tex
\begin{tikzpicture}[->,>=stealth',shorten >=1pt,auto,node distance=2.5cm, initial text=]\scriptsize
  \tikzstyle{every state}=[fill=none,draw=black,text=black,inner sep=1.5pt, minimum size=12pt,thick,scale=0.7]

  \node[state, initial above] (i)     {$\mathit{init}$};
  \node[state, accepting] (t) [below of=i,yshift=.7cm] {$\mathit{iter}$}; 
  	\node[state] (l) [right of=d,xshift=.5cm] {$\mathit{loop}$};
  	\node[state] (s) [right of=i,xshift=.5cm]    {$\mathit{sink}$};
 
 	\path (i) edge  node[below,yshift=-.15cm,sloped] {$c' = 123$} (t);
	\path (t) edge [bend left=10] node[above] {$\beta$} (l);
 	\path (t) edge node[above,sloped] {$c < -200$} (s);
  	\path (l) edge [bend left=10] node[below] {$y \leq 0 \land c' = c$} (t);
 	\path (l) edge[loop above]  node[above] {$\alpha \land c' = c$} (l);
  	\path (s) edge[loop above]  node[left, near start] {$\top$} (s);
\end{tikzpicture}

%% file: games.tex
\paragraph{First-order logic.}
We consider a set $\vars$ of all variables.
For $X \subseteq \vars$, an \emph{assignment to} $X$ is a
function $\assignment: X \to \values$ where $\values$ is the set of all values.
We define $\assignments{X}$ to be the assignments to $X$.
$\assignment_1\uplus \assignment_2$ denotes the combination of two assignments $\assignment_1, \assignment_2$ to disjoint variables.
$X'$  is the \emph{primed version of $X$} such that
$X' : = \{x' \mid x \in X\} \subseteq \vars \setminus X$.
If $\assignment \in \assignments{X}$,  we define $\assignment' \in \assignments{X'}$ as $\assignment'(x') = \assignment(x)$ for all $x \in X$.
Given $\assignment_1,\assignment_2 \in \assignments{X}$,
we define $\langle \assignment_1, \assignment_2 \rangle := \assignment_1 \uplus \assignment_2'$.
Let $\FOLX$ and $\QFX$ be the sets of first-order and quantifier-free formulas, respectively.
For $\alpha \in \FOLX$ and $X = \{x_1,\ldots,x_n\} \subseteq \vars$,  we write  $\alpha(X)$ when the free variables of $\alpha$ are a subset of $X$.
We denote with $\FOL{X}$ ($\QF{X}$) the formulas (quantifier-free formulas) whose free variables belong to $X$.
We use $Q X.\alpha$ as a shortcut for $Q x_1.\ldots Q x_n.\alpha$ for $Q \in \{\exists, \forall\}$.
The formula $\alpha[x_1 \mapsto y_1, \dots, x_n \mapsto y_n]$ is obtained from $\alpha$ by replacing all $x_i$  simultaneously by variables $y_i$.
We denote by $\assignment \FOLentailsT{T} \alpha$ entailment of $\alpha$ by $\assignment$ in the first-order theory $T$ for $\alpha \in \FOL{X}$ and $\assignment \in \assignments{X}$.
We assume a fixed theory $T$ and define $\sema{\alpha(X)} := \{\assignment \in \assignments{X} \mid \assignment \FOLentailsT{T}  \alpha\}$.
Independent of the theory, we use uninterpreted functions as syntactic place-holders.
In their presence entailment is undefined.
We call them uninterpreted constants if they have arity zero.
$\alpha[c_1 \mapsto x_1, \dots, c_n \mapsto x_n]$ is obtained from $\alpha$ by replacing all uninterpreted constant symbols $c_i$ simultaneously by $x_i$.
For variables $Y= \{y_1, \dots, y_m\}$, uninterpreted functions symbols $f_i$ with arity $m$ over $Y$, and terms $\theta_i$ over $Y$,
$\alpha\llparenthesis f_1 \mapsto \theta_1, \dots, f_n \mapsto \theta_n \rrparenthesis$ is obtained by replacing all instances $f_i(t_1, \dots, t_m)$ in $\alpha$ by $\theta_i$ where all variables $y_j$ are replaced by $t_j$.

\paragraph{Two-Player Turn-Based Games.}

An \emph{arena} is a tuple $G = (V,V_\env,V_\sys,\tau)$ where $V = V_\env \uplus V_\sys$ are the vertices,  partitioned between the environment ($\env$) and the system player ($\sys$), and $\tau \subseteq (V_\env \times V_\sys) \cup (V_\sys \times V_\env)$ is the transition relation.
A \emph{play} in $G$ is a sequence $\xi \in V^\omega$ where $(\xi[i],\xi[i+1])\in\tau$ for all $i \in \Nat$.
A \emph{strategy for player~$p$} is a function
$\sigma: V^*V_{p} \to V$ where $\sigma(\xi\cdot v) = v'$ implies $(v,v') \in \tau$.  $\str{p}{G}$ are the strategies for $p$ in $G$.
A play $\xi$ is consistent with $\sigma$ if $\xi[i+1] = \sigma(\xi[0,i])$ for every $i \in \Nat$ where $\xi[i] \in V_p$.
$\plays_G(v,\sigma)$ is the set of all plays in $G$ starting in $v$ and consistent with $\sigma$.
The pair $(G,\Omega)$ is a \emph{two-player turn-based game} where $\Omega \subseteq V^\omega$ is called the \emph{winning condition} for $\sys$.
A sequence $\xi \in V^\omega$ is \emph{winning for $\sys$} if and only if $\xi \in \Omega$,  and is \emph{winning for $\env$} otherwise.
A strategy $\sigma$ of player $p$ is \emph{winning for $p$ from $v$} if every play in $\plays_G(v,\sigma)$ is winning for player~$p$.

\subsection{Representation and Methods for Solving Infinite-State Games}

We follow the formalization in~\cite{HeimD25} and represent infinite arenas with \emph{symbolic game structures}.
Those describe the interaction between an environment selecting values for the input variables $\inputs$ and a reactive system controlling the program variables $\progvars$.
Formally, a \emph{symbolic game structure} is a tuple $(L, \linit,\inputs, \progvars, \dom, \delta)$ where
$L$ is a finite set of \emph{locations,}
$\linit \in L$ is the \emph{initial location},
$\inputs \subseteq \vars$ is a finite set of \emph{input variables},
$\progvars \subseteq \vars$ is a finite set of \emph{program variables},
$\mathit{dom}: L \mapsto \QF{\progvars}$ is the \emph{domain of the states},
and $\delta: L \times L \mapsto \QF{\progvars \cup \inputs \cup \progvars'}$ is the \emph{transition relation}.
For $l \in L$ and  $\assmt{x} \in \assignments{\progvars}$ with $\assmt{x} \FOLentailsT{T} \mathit{dom}(l)$, $\assmt{i} \in \assignments{\inputs}$ is a \emph{valid input} if there exists some $l' \in L$ and $\assmt{x}' \in \assignments{\progvars'}$ such that $\assmt{x}' \FOLentailsT{T} \mathit{dom}(l')$ and $\assmt{x} \uplus \assmt{i} \uplus \assmt{x}' \FOLentailsT{T} \delta(l, l')$.
$\delta$ is required to be \emph{non-blocking}, i.e., for all $l \in L$ and  $\assmt{x} \in \assignments{\progvars}$ with $\assmt{x} \FOLentailsT{T} \mathit{dom}(l)$ there exists some valid input.

\paragraph{Semantics.}

A symbolic game structure $\symgame$ represents a possibly infinite arena $\sema{\symgame} = (\states, \states_\env,\states_\sys,\tau)$.
$\states_\env$ consists of pairs $(l,\assmt{x})$ of location $l \in L$ and assignment to the program variables $\assmt{x} \in \assignments{\progvars}$ such that the assignment $\assmt{x}$ satisfies the condition $\mathit{dom}(l)$, i.e.\ $\states_\env := \{ (l,\assmt{x}) \in L \times \assignments{\progvars} \mid \assmt{x} \FOLentailsT{T} \mathit{dom}(l)\}$.
Those vertices are also called \emph{states}.
From such a state,  $\env$ selects a valid input $\assmt{i} \in \assignments{\inputs}$.
Hence, $\states_\sys$ are pairs of states and respective valid inputs, i.e.\
$\states_\sys :=\{ ((l,\assmt{x}),\assmt{i}) \in \states_\env\times \assignments{\inputs}\mid \validinput(l, \assmt{x}, \assmt{i}) \}$.
Then, $\sys$ selects a next state  $(l', \assmt{x}')$ with a new assignment to the program variables and next location $l' \in L$ such that $\assmt{x'}  \FOLentailsT{T}  \dom(l')$ and $\assmt{x}\uplus \assmt{i} \uplus \assmt{v}' \FOLentailsT{T} \delta(l,l')$.
The transition relation $\tau \subseteq (\states_\env \times \states_\sys) \cup (\states_\sys \times \states_\env)$ describes these interactions.
Formally, $\tau$ is the largest relation such that
$(((l,\assmt{x}), \assmt{i}),(l',\assmt{v})) \in \tau$ implies that $\assmt{x} \uplus\assmt{i} \uplus \assmt{v}' \FOLentailsT{T} \delta(l,l')$ holds.

\paragraph{Winning Condition, Realizability and Synthesis Problem.}

We consider so called location-based winning conditions defined in terms of the infinite sequence of locations $\Lambda \subseteq L^\omega$ visited in a play.
Concretely,  we  consider location-based parity conditions defined by a coloring function $\lambda: L \to  \Nat$ that consist of $l_0l_1  \ldots \in L^\omega$ where the largest number occurring infinitely often in $\lambda(l_0)\lambda(l_1)$ is odd.
These are sufficient to describe all $\omega$-regular temporal properties.
The \emph{realizability problem} is to compute whether there exists a strategy $\sigma$ for $\sys$ such that for all $\assmt{x}  \FOLentailsT{T}  \dom(\linit)$, $\sigma$ is winning for $\sys$ from $(\linit, \assmt{x})$.
The \emph{synthesis problem} is to additionally compute some representation for $\sigma$.

\SetAlgoSkip{0}
\SetKwComment{Comment}{/* }{ */}
\begin{algorithm}[t!]
\SetAlgoVlined
\SetKwProg{Fn}{function}{}{}
\DontPrintSemicolon
\Fn{\textsc{AttractorAccel}(
	 $\symgame$,
	 $p \in \{\sys,\env\}$,
	 $\mathit{target} \in \symstates$)}{		
	 \nl $a^0$ := $\lambda l.~\bot$;
         $a^1$ := $\mathit{target}$\;
     \nl \For{$n=1,2,\ldots$}{
		\nl \lIf{$a^{n} \equiv_{T} a^{n-1}$}{\Return $a^n$}
        $a^n(l_a)$ := $a^n(l_a) \lor \textsc{Accelerate}(\symgame, p, a^n, l_a)$ for some $l_a \in L$
		\nl $a^{n+1} := a^n \lor \cpre{\mathcal G,p}{a^n}$\label{line:cpre}\;
}
}
\caption{Attractor Computation with Acceleration from \cite{HeimD24}.}
\label{alg:attractor-with-accel}
\end{algorithm}

\paragraph{Attractors.}
A key operation for solving symbolic games is \emph{symbolic attractor computation} which is leveraged to solve general symbolic games~\cite{HeimD24}.
Hence, we focus on the \textbf{problem of computing attractors in symbolic games}.
Intuitively, an attractor is the set of states from which a given player~$p$ can enforce reaching a set of target states no matter what the other player does.
Formally, the \emph{player-$p$ attractor for $R \subseteq \states$} in a symbolic game structure $\symgame$ is
$\mathit{Attr}_{\symgame,p}(R) :=
\{s \in \states \mid \exists \sigma \in \str{p}{\sema{\mathcal G}}.\forall \pi \in \plays_{\sema{\symgame}}(s,\sigma).
\exists n.\;\pi_n\in R\}.$
We represent infinite sets of states using the \emph{symbolic states} $\symstates: L \to \FOL{\progvars}$.
For $d \in \symstates$, the concrete states of $d$ are $\sema{d} := \{(l, \assmt{x}) \mid \assmt{x} \FOLentailsT{T} d(l) \land  \assmt{x} \FOLentailsT{T} \dom(l)\}$.
Symbolic attractors are computed using the \emph{symbolic enforceable predecessor operator} $\cpreX{\symgame,p}: \symstates \to \symstates$. $\cpre{\symgame,p}{d}$ represents the states from which player~$p$ can enforce reaching $\sema{d}$ in one step in $\symgame$ (i.e. one move by each player). Formally,
{\footnotesize
\[
\begin{array}{ll}
\cpre{\symgame,\sys}{d}(l) :=&
\dom(l) \land
\forall \inputs. \validinput(l) \to
\exists \progvars'. \bigvee_{l' \in L}\delta(l,l') \land \circ \dom(l') \land \circ d(l')\\
\cpre{\symgame,\env}{d}(l) : =&
\dom(l) \land
\exists \inputs. \validinput(l) \land
\forall \progvars'. \bigwedge_{l' \in L}\delta(l,l') \land \circ\dom(l') \to \circ d(l')
\end{array}
\]
}
where $\circ \varphi := \varphi[\progvars \mapsto \progvars']$ and $\validinput(l):= \exists \progvars'\bigvee_{l' \in L}. \delta(l,l') \land \circ\dom(l')$.

Using  $\cpreX{\symgame,p}$,  the player-$p$ attractor can be computed as a fixpoint as shown in \Cref{alg:attractor-with-accel}.
For games with infinite state-spaces,  an iterative attractor computation might not terminate.
Therefore, \cite{HeimD24} introduced \emph{attractor acceleration}, which extends the computed attractor
to help finding the fixpoint as shown in \Cref{alg:attractor-with-accel}.
We explain how attractor acceleration works in the next section.

\paragraph{Theories.} 

The above notions are independent of the theory $T$.
The game-solving methods rely on satisfiability and, sometimes, quantifier elimination, although the latter can often be approximated soundly with the former.
In practice, quantifier elimination is crucial to simplify terms.
We focus on linear arithmetic.

%% file: lemmas.tex
The iterative symbolic computation of attractors in infinite-state games is not guaranteed to terminate.
As discussed in~\Cref{sec:intro}, \cite{HeimD24} introduced a method that aims to alleviate this by \emph{accelerating the attractor computation}. This is done by computing a form of inductive statements that establish for some (infinite) set of states the existence of a strategy for a player to enforce reachability of a given target. In~\cite{HeimD24}, these inductive statements are formalized using the notion of \emph{acceleration lemmas}, which intuitively state that:
\textit{``If starting from a set of conclusion states, a player can always enforce some step relation between states, then eventually some base set is reached.''}.
\cite{HeimD24} computes acceleration lemmas by representing them via uninterpreted functions that are used in the symbolic computation. 
In the process, a set of constraints over these functions is constructed and must be solved to determine if acceleration can be applied and derive the respective strategy. 
A constraint solver then searches for an applicable acceleration lemma.
As the constraints encode the existence of a strategy in the game, they quickly get quite complex, leading to scalability issues. 

In this paper, we propose a new approach for computing acceleration arguments.
In contrast to the ``one-shot'' approach of~\cite{HeimD24}, our method constructs such arguments in a modular fashion by composing them from simple ones and performs the search for the argument itself.
To enable composition, we introduce a novel notion, called
\emph{generalized acceleration lemmas (GALs)}, extending acceleration lemmas. Intuitively, a GAL states that
\textit{``If starting from a set of conclusion states, a player can \textbf{infinitely-often} enforce some step relation between states \textbf{and otherwise not lose progress}, then eventually some base set is reached.''}
Lifting the condition that the step relation is always enforced is crucial for combining arguments. It allows making progress in one part of the argument without having to make progress in another part (while not losing progress there). This is useful, for instance, for lexicographic arguments like the one in~\Cref{ex:intro-reach}, where $y$ remains the same while $x$ decreases.

A GAL, formally introduced below, is a tuple $(\base, \stay, \step, \conc)$ of \FOLX\ formulas.
The \emph{base condition} $\base$ characterizes a set of target states.
The \emph{step relation} $\step$ and the \emph{stay relation} are relations between states.
The \emph{conclusion} $\conc$ characterizes states from which every sequence of states in which each pair of consecutive states conforms to either $\step$ or $\stay$ will necessarily reach the target set $\base$, if it conforms to $\step$ infinitely often.
Thus, the relation $\step$ captures a ranking argument for establishing the reachability of $\base$ starting from $\conc$, while $\stay$ captures preservation of progress.
As a simple example, consider the GAL $(y \leq 0, y' \leq y, y' < y, \top)$ where $y$ is an integer variable, which captures the fact that any sequence where $y$ decreases infinitely often and otherwise does not increase must eventually reach a state where $y$ is non-positive.

\begin{definition}[Generalized Acceleration Lemma]\label{def:gen-lemma}
A \emph{generalized acceleration lemma (GAL)} is a tuple
$(\base, \stay, \step, \conc)$
of first-order formulas $\base, \conc \in \FOL{V}$ and $\stay, \step \in \FOL{V \cup V'}$
 for some $V \subseteq \vars$, where
\begin{enumerate}
\item[(I)]
For every sequence
$\alpha \in \assignments{V}^\omega$, if
$\alpha[0] \FOLentailsT{T} \conc$,
and
\begin{enumerate}
    \item[(a)]   for all $i \in \Nat$, $\combine{\alpha[i]}{\alpha[i+1]} \FOLentailsT{T} \step \lor \stay$ and
    \item[(b)]   for all $i \in \Nat$ there exists $j \geq i$ such that $\combine{\alpha[j]}{\alpha[j+1]} \FOLentailsT{T} \step$,
\end{enumerate}
then there exists some $k \in \Nat$ such that $\alpha[k] \FOLentailsT{T} \base$,

\item[(II)] and,
for all $\assignment, \assignment' \in \assignments{V}$ with
$\assignment \FOLentailsT{T} \conc$ and
$\combine{\assignment}{\assignment'} \FOLentailsT{T} \step \vee \stay$ we have that
$\assignment' \FOLentailsT{T} \conc$ holds.
\end{enumerate}
\end{definition}

\subsection{Using (Generalized) Acceleration Lemmas (adapted from~\cite{HeimD24})}
Before we present our novel method for the computation of GALs, we first explain how (generalized) acceleration lemmas are used to accelerate attractor computation.
In Algorithm~\ref{alg:acceleration}, we re-frame the procedure from~\cite{HeimD24}, in that we clearly distinguish the two parts of the lemma computation: (1) the construction of a candidate lemma in line~\ref{line:pick-gal} of the algorithm, and (2) checking that the step of the lemma can actually be enforced by the respective player in the game, expressed using the condition constructed in line~\ref{line:step-cond}.

In~\cite{HeimD24}, the two aspects are heavily intertwined. There, \textsc{pick} simply introduces uninterpreted functions representing the lemma that is passed to the symbolic procedure for generating and verifying the applicability conditions. If the latter requires acceleration, further lemmas and constraints are accumulated.

In contrast, the approach we describe in~\Cref{sec:gal-rules} and \Cref{sec:gal-computation} analyzes the target of the acceleration and the game
to construct a concrete GAL on which the enforcement condition is then checked.
An important feature of our GAL construction is that it uses the structure of the target and the game's transition relation to derive complex GALs from simpler ones using the composition operations presented in~\Cref{sec:gal-rules}.
As we demonstrate in \Cref{sec:experiments}, this significantly improves performance when complex acceleration arguments are needed.\looseness=-1

In more detail, Algorithm~\ref{alg:acceleration} shows how acceleration is computed.
The procedure accelerates iterative attractor computations based on sub-games that contain a loop at a location $l$ in the respective location graph.
A lemma must satisfy two conditions:
First, $\base$ should be part of the target player~$p$ tries to enforce (first conjunct in line \ref{line:conds-check}).
The second conjunct in line \ref{line:conds-check} requires that every state $s$ that satisfies
$\conc \land \neg \base$ also satisfies the loop-step condition $\psi$ in \ref{line:step-cond}, meaning that it should be possible to enforce from $s$ in location $l$ to come back to $l$ with a state $s'$ such that $s$ and $s'$ satisfy $\step$.
This condition is computed by \textsc{LoopStep}.\looseness=-1

First, \textsc{LoopStep} constructs a so-called \emph{loop game}~\cite{HeimD24} from location $l$ to a copy of itself, obtained by ``braking '' the loop in $l$.
$\loopgame(\symgame, l, \locend)$ is the symbolic game structure obtained by adding a new location $\locend$ and redirecting all edges in $\symgame$ with target location $l$ to the new location $\locend$.
The following formal definition has been adapted from~\cite{HeimD24}.

\medskip
\noindent
\textbf{Defintion (Loop Game).}
\textit{%
Given a symbolic game structure \\ $\symgame  = (L, \linit,\inputs, \progvars, \dom, \delta)$, a location $\locsplit \in L$, and a fresh location $\locend \not\in L$,
the \emph{loop game}
is the symbolic game structure
$\loopgame(\symgame,  \locsplit, \locend) := (L\cup \{\locend\},\inputs, \progvars,\dom',\delta')$, where
$\dom': = \dom \cup \{\locend \mapsto \dom(\locsplit)\}$, and
$\delta'(l_1,l_2) := \delta(l_1,l_2)$ if $l_1,l_2 \neq \locend$,
$\delta'(l_1,\locend):= \delta(l_1,\locsplit)$ if $l_1 \neq \locend$, and
$\delta'(\locend,\locend):= \bigwedge_{x \in \progvars} x' = x$.%
}
\medskip

Then, \textsc{LoopStep} computes the attractor (or an under-approximation of it) to reach $\step$ (with swapped variables, since we are doing backward computation) in the copied location of the loop game (or reach $\mathit{target}$ directly).
For more intuition, we refer the reader to Example 7.5 from~\cite{HeimD24}.

\SetAlgoSkip{0}
\SetKwComment{Comment}{/* }{ */}
\begin{algorithm}[t!]
\SetAlgoVlined
\SetKwProg{Fn}{function}{}{}
\DontPrintSemicolon
\Fn{\textsc{Accelerate}(
	 $\symgame$,
	 $p \in \{\sys, \env\}$,
	 $\mathit{target} \in \symstates$,
     $l \in L$)}{		
     \textsc{pick} $(\base, \stay, \step, \conc) \in \mathit{GALs}$\label{line:pick-gal}\;
     $\psi : = \textsc{LoopStep}(\symgame, p, \mathit{target}, l, \step)$\label{line:step-cond}\;
     \lIf{$(\forall \progvars.~\base \land \dom(l) \to \mathit{target}(l))$\textbf{ and }$(\forall \progvars.~\conc \land \lnot \base \land \dom(l) \to \psi)$\label{line:conds-check}}{
            \Return $\conc \land \dom(l)$}
     \Return $\bot$\;
}
\Fn{\textsc{LoopStep}(
	 $\symgame$,
	 $p \in \{\sys, \env\}$,
	 $\mathit{target} \in \symstates$,
     $l \in L$,
     $\step \in \FOL{\progvars}$)}{
     $\symgame_\mathit{Loop}$ := $\loopgame(\symgame, l, \locend)$ for $\locend \not\in L$\;
     Let $E := \{ e_x \mid x \in \progvars \}$ be fresh uninterpreted constants\;
     $\mathit{target}_\mathit{loop}$ := $\lambda l'. \textsc{if } l' = \locend \textsc{ then } \step[\progvars \mapsto E, \progvars' \mapsto \progvars] \textsc{ else } \mathit{target}(l')$\;
     $d_\mathit{Loop}$ := $\textsc{Attractor}(\symgame_\mathit{Loop}, p, \mathit{target}_\mathit{loop})$ \Comment{can under-approximate}\nl
     \Return $d_\mathit{Loop}(l)[E \mapsto \progvars]$\;
}

\caption{Acceleration procedure re-framed from \cite{HeimD24}.}\label{alg:acceleration}
\end{algorithm}

\subsection{Composition Operations for Generalized Acceleration Lemmas}\label{sec:gal-rules}
We now turn to the first key ingredient of our method for the synthesis of GALs, the operations for composing simpler GALs into more complex ones.
The proofs of all statements in the rest of the section can be found in Appendix~\ref{app:proofs}.

Two key composition operators are \emph{intersection} and \emph{lexicographic union}.
The first one allows for reasoning about progress towards conjunctions of targets in a modular way. As established in the next lemma, if we start in the intersection of the conclusions of two GALs, we can reach the intersection of the respective bases by making progress in one of the GALs while not losing progress in the other one. \looseness=-1

\begin{lemma}[Intersection of GALs]\label{lem:gal-intersect}
Let $(\base_0, \stay_0, \step_0, \conc_0)$ and \\$(\base_1, \stay_1, \step_1, \conc_1)$ be GALs over the same variables $V$. Then, for \\
$\step := \mathit{stayBase} \land \bigvee_{i \in \{0, 1\}}(\step_i \land \lnot\base_i \land \stay_{1 - i})$ and \\
$\mathit{stayBase} := \bigwedge_{i \in \{0,1\}}((\base_i \land \lnot \base_{1 - i}) \to \base_i[\progvars \mapsto \progvars'])$,
the tuple\\ $(\base_0 \land \base_1, \stay_0 \land \stay_1 \land \mathit{stayBase}, \step, \conc_0 \land \conc_1)$ is also a GAL.
\end{lemma}
Note that $\step$ ensures that we do not stop making progress in one of the GALs after reaching the base of the other, by excluding the bases from doing a step.

The second combination, akin to lexicographic combinations of ranking functions, allows us to combine two GALs via union.
The union combination is lexicographic, i.e.\ if we make progress on the first GAL, this is fine regardless of the other one. If instead we only progress on the second one, we should not lose progress in the first one.
The last condition ensures that it is not possible to jump back and forth between the two GALs to cancel out the progress made.

\begin{lemma}[Lexicographic-Union of GALs]\label{lem:gal-union}
Let $(\base_0, \stay_0, \step_0, \conc_0)$ and $(\base_1, \stay_1, \step_1, \conc_1)$ be GALs over the same variables $V$. Then, for
$\step := (\conc_0 \land \step_0 ) \lor (\conc_1 \land \step_1 \land \stay_0)$,
the tuple $(\base_0 \lor \base_1, \stay_0 \land \stay_1, \step, \conc_0 \lor \conc_1)$ is also a GAL.
\end{lemma}

\begin{example}\label{ex:lemma-running}
These two operations allow us to build up the enforcement argument for \Cref{ex:intro-reach} from individual ones for each of the two variables.
The intersection $\mathit{lem}_{x,y}$ of $(y \leq 0,y' \leq y,y' < y,\top)$ and $( x \leq 0, x' \leq x,x'< x,\top)$ ensures that $y \leq 0 \land x \leq 0$ is reached if each of $x$ and $y$ can be decreased while the other one is not increasing. However, this cannot be enforced in the game in \Cref{fig:example-reach}, since by always setting $b$ to true in $\mathit{loop}$ and picking larger and larger values for $i$, the environment can prevent progress on $x$.
If, however, we take the lexicographic union with first element
$\mathit{lem}_{y} : = (y \leq 0, y' \leq y,y'<y,\top)$ and second element $\mathit{lem}_{x,y}$ we obtain a GAL that states that $y \leq 0 \lor (y \leq 0 \land x \leq 0)$ can be reached from any state if the lexicographic step relation $(y' < y) \lor (\step_{x,y} \land y' \leq y)$ can be enforced. This is indeed possible, as explained informally in \Cref{ex:intro-reach}. This GAL is the one automatically constructed by our implementation and is sufficient for accelerating the attractor computation for the game in \Cref{fig:example-reach}.
\end{example}

The GAL constructed in the previous example requires the introduction of the sub-lemma about $x$, which enables the main argument on $y$.
It is sometimes easier to find such arguments in the form of a ``chain'' of two GALs by analysis of the transition relations in the game structure.
More concretely, we can chain two GALs by extending the step relation of the first GAL by the possibility to perform the step of the second GAL until it reaches its base. This can then be used as an additional condition enabling a step in the first GAL.
However, theoretically this is subsumed by applying Lemmas~\ref{lem:gal-intersect} and \ref{lem:gal-union} to $\varphi \equiv \varphi \lor (\varphi \land \psi)$.
\begin{lemma}[Chaining of GALs]\label{lem:gal-chain}
Let $(\base_0, \stay_0, \step_0, \conc_0)$ and\\ $(\base_1, \stay_1, \step_1, \conc_1)$ be GALs over the same variables $V$. Then for
$\step := \step_0 \lor (\conc_1 \land \lnot \base_1 \land \step_1 \land \stay_0)$ and 
$\stay := \stay_0 \land \stay_1 \land (\base_1 \to \base_1[\progvars \mapsto \progvars'])$
the tuple $(\base_0, \stay, \step, \conc_0)$ is also a GAL.
\end{lemma}

In complex games, it is useful to restrict a GAL to only take effect in a specific sub-part of the state space by adding an invariant as follows.
\begin{lemma}[Invariant Strengthening of GALs (extended from~\cite{HeimD24})]\label{lem:gal-inv}
Let $(\base, \stay, \step, \conc)$ be a GAL over $V$ and $\inv \in \FOL{V}$.
Then for $\inv' := \inv[V \mapsto V']$,
$(\base \land \inv, \stay \land \inv', \step \land \inv', \conc \land \inv)$ is also a GAL.\looseness=-1
\end{lemma}

\subsection{Computation of Generalized Acceleration Lemmas}\label{sec:gal-computation}

In the rest of this section, we present our method for computing GALs that are possibly useful to reach a given $\mathit{target} \in \symstates$ in a location $l \in L$. We use the operations from the previous subsection to compose GALs as the building blocks forming the base layer for applying the combinations. We use GALs whose components are inequalities between affine terms, as formalized in the next lemma.

\begin{lemma}[Inequality Base GALs]\label{lem:gal-base}
Let $t$ be a linear term over numeric variables $V$ and
let $a \in \Real \cup \{-\infty\}$, $b \in \Real \cup \{\infty\}$ with $a \leq b$ be bounds.
Then, $(\base, \stay, \step, \conc)$ is a GAL for $t' := t[V \mapsto V']$ and $\epsilon > 0$,
\begin{itemize}
    \item   $\base := a \leq t \leq b$, $\conc := \top$,
    \item   $\stay := (a \leq t' \leq b) \lor (t < a \land t \leq t' \leq b) \lor (t > b \land t \geq t' \geq a)$, and
    \item   $\step := (a \leq t' \leq b) \lor (t < a \land t + \epsilon \leq t' \leq b) \lor (t > b \land t - \epsilon \geq t' \geq a).$
\end{itemize}
\end{lemma}
Note that there also exist variants of the above GAL with strict inequalities, but we omit those for the brevity of the presentation.
Also, \Cref{lem:gal-base} is specific to linear arithmetic, and other theories would need analogous lemmas.

Starting from inequality base GALs, our method proceeds in two phases.
First, we consider the formula $\mathit{target}(l)$ in order to derive suitable base GALs, strengthen them with invariants, and build combinations thereof.
Next, we consider the given game structure $\mathcal G$ and the conditions that need to be satisfied to enforce the step relation in $\mathcal G$. We now describe the two phases in detail.

We begin by rewriting $\mathit{target}(l)$ into a disjunction of polyhedra conjuncted with (non-linear) terms (which is always possible), i.e., into the form
$\mathit{target}(l) \equiv \bigvee_{i = 1, \dots, n} A_i \mathbf{x} \leq c_i \land \psi_i(\progvars)$
for matrices $A_i$, constant vectors $c_i$, a vector $\mathbf{x}$ for the numeric variables in $\progvars$ and remaining constraints $\psi_i$ for all variables.
For each disjunct, we select a subset of the inequalities of the polyhedra to get GALs by \Cref{lem:gal-base}, which we combine using \Cref{lem:gal-intersect}.
We then add the remaining inequalities and the non-linear terms as invariants using \Cref{lem:gal-inv}.
The disjuncts (or a subset of those) can then be combined in different orders by \Cref{lem:gal-union}.
These steps are carried out in line~\ref{line:pick-gal-target} of our method for constructing GALs shown in \Cref{alg:computation-of-lemmas}.
As there are multiple combinations, this provides a well-defined space of lemmas to search through based on the $\mathit{target}(l)$ that we want to reach.

\begin{example}
Suppose that $\mathit{target}(l) = (q \land 3 x + 2y \leq z \land z \leq 4x \lor \neg q \land 2x = y)$ for Boolean input $q$ and numerical variables $x,y$ and $z$.
One option is to pick $3 x + 2y \leq z$ from the first disjuntct and $y \leq 2x$ from the second one, and use each of them as the $\base$ for a base GAL as per \Cref{lem:gal-base}. After that, applying \Cref{lem:gal-inv}, we strengthen each of them with an invariant consisting of the Boolean term and the remaining inequality from the respective disjunct. Finally, we can combine the resulting GALs using \Cref{lem:gal-union}.
\end{example}

Next, we account for the need to enforce the step relation in the game, which also allows us to generate sub-arguments for chain combinations via \Cref{lem:gal-chain}.
To this end, in \Cref{alg:computation-of-lemmas} we consider the \emph{enforcement condition for acceleration} lemmas as computed by \textsc{LoopStep} (shown in
\Cref{alg:acceleration}) and use it to generate the intermediate arguments.
More precisely, $\mathit{pre}$ in line~\ref{line:pre-in-get-lemma} is the condition that enables player~$p$ to enforce the step relation \emph{once}.
If this condition is not implied by the GAL's precondition (that is, the check in line~\ref{line:check-pre} fails), we attempt to fulfill it by either making it an invariant or by computing a sub-lemma that establishes this condition as a sub-argument.
Note that we might want to iterate this process, as, e.g., strengthening the lemma with an invariant makes the step condition harder to apply and might need even a stronger invariant.
While this process is not guaranteed to succeed with an applicable lemma, it provides a robust method for computing GALs, as demonstrated by our experimental evaluation.

Note that $\textsc{Recurse?}$ and $\textsc{Iterate?}$ are heuristics that determine the space of GALs  we search through.
In practice, we found it effective to start with a few iterations and recursions, gradually increasing them if acceleration fails.

\SetAlgoSkip{0}
\SetKwComment{Comment}{/* }{ */}
\begin{algorithm}[t!]
\SetAlgoVlined
    \SetKwProg{Fn}{function}{}{}
    \DontPrintSemicolon
    \Fn{\textsc{GetGAL}(
		 $\symgame$,
    	 $p \in \{\sys,\env\}$,
		 $l \in L$,
		 $\mathit{target} \in \symstates$)}{
    	 $(\base, \stay, \step, \conc)$ := Get lemma from $\mathit{target}$ using Lemmas~\ref{lem:gal-base}, \ref{lem:gal-inv}, \ref{lem:gal-union}, \ref{lem:gal-intersect}\label{line:pick-gal-target}\;
         \While{$\textsc{Iterate?}$}{
    	 $\mathit{pre} :=\textsc{LoopStep}(\symgame, p, \mathit{target}, l,\step)$\label{line:pre-in-get-lemma}\;
         \If{$\forall \progvars.~\conc \land \lnot \base \land \dom(l) \to \mathit{pre}$\label{line:check-pre}}{
            \Return $(\base, \stay, \step, \conc)$\;
         }
    	 \eIf{\textsc{Recurse?}}{
            $\mathit{subGAL}$ := $\textsc{GetGAL}(\symgame, p, l, \lambda l'. \textsc{if } l = l' \textsc{ then } \mathit{pre} \textsc{ else } \bot)$\;
            Chain $\mathit{subGAL}$ to $(\base, \stay, \step, \conc)$ using Lemma~\ref{lem:gal-chain}\;
    	 }{
            Add $\mathit{pre}$ as invariant to $(\base, \stay, \step, \conc)$ with Lemma~\ref{lem:gal-inv}\;
         }
        }
	}
\caption{Computation of generalized acceleration lemmas.}
\label{alg:computation-of-lemmas}
\end{algorithm}

%% file: summaries.tex
In this section, we define enforcement summaries and demonstrate their use and computation.
Intuitively, an enforcement summary is a \emph{parametrized formula} $\varphi$ that states that player~$p$ can enforce to reach a symbolic state $d$ starting from a location $l_s$ and assignment for $\progvars$ that $\varphi$ describes when its parameters are instantiated by $d$.
To insert values of a symbolic state into a formula, we represent the parameters as uninterpreted function symbols. 
However, those are merely syntactic placeholders, and we perform no computation on them.
In most cases, it is not possible to fully characterize in FOL all states from which any arbitrary symbolic state $d$ is enforceable by a given player. 
Hence, enforcement summaries have the following characteristics:
First, they can \emph{under-approximate} the starting set from which $d$ is enforceable.
Second, they do not apply to all possible $d$ but can be \emph{restricted to a subset} of $\symstates$.
Formally, that is:

\begin{definition}[Enforcement Summary]\label{def:summary}
Let $\symgame = (L, \linit,\inputs, \progvars, \dom, \delta)$ be a symbolic game structure.
An \emph{enforcement summary} in $\symgame$ is a tuple 
$(p,l_s,\varphi, D)$ where 
$p \in \{\env, \sys\}$, 
$l_s \in L$ is the support location,
$D \subseteq \symstates$ is the support set, and
$\varphi \in \FOL{\progvars}$ is the summary statement with 
uninterpreted function symbols $\nextp_l$ over $\progvars$ for all $l \in L$ 
such that for every $d \in D$ we have
    \[\sema{ \varphi\llparenthesis\nextp_l \mapsto d(l) \mid l \in L\rrparenthesis} \subseteq \mathit{Attr}_{\symgame,p}(\sema{d})(l_s). \]
\end{definition}

For a finite characterization of $D$, we use template formulas over state and \emph{meta-variables}.
Those describe all symbolic states for which meta-values exist such that the template is a subset of the symbolic state.
More precisely, a \emph{set of meta-variables} $\metavars \subseteq \vars \setminus \progvars$ and a template $\tau: L \to \FOL{\progvars \cup \metavars}$ define the support set 
$D = \left\{ d \in \symstates \mid \exists \metavars. \forall l \in L. \forall \progvars. \tau(l) \to d(l)~\text{holds} \right\}$.

\paragraph{Usage.}
The application of enforcement summaries follows \Cref{def:summary}. 
During attractor computation for player~$p$ in $\symgame$, if for a summary $(p,l_s,\varphi, D)$ the current attractor $a$ is in $D$, we extend $a(l_s)$ by $\varphi$ with the uninterpreted functions substituted by the current values of $a$. By  \Cref{def:summary}, we have that the added set is indeed subset of  $\mathit{Attr}_{\symgame,p}(\sema{\mathit{target}})(l_s)$.
\Cref{alg:use-summaries} formalizes this.

\SetAlgoSkip{0}
\SetKwComment{Comment}{/* }{ */}
\begin{algorithm}[t!]
\SetAlgoVlined
    \SetKwProg{Fn}{function}{}{}
    \DontPrintSemicolon
    \Fn{\textsc{AttractorSumm}(
		 $\mathcal G$,  
    	 $\mathit{p} \in \{\sys,\env\}$,  
		 $\mathit{target} \in \symstates$, 
		 \changed{$\Summaries$})}{		
         $\dots$ \;
         \setcounter{AlgoLine}{6}
		 \lIf{$a^{n} \equiv_{T} a^{n-1}$}{\Return $a^n$}
            \changed{
            \ForEach{$(p, l_s, \varphi, D) \in \Summaries$ for $\symgame$}{
            \If{$a^n \in D$\Comment{Checked by validity query for representation.}}{
            $a^n(l_s) := a^n(l_s) \;\lor\; \varphi \llparenthesis \nextp_{l} \mapsto a^n(l) \mid l \in L \rrparenthesis$\label{line:use-summary}
            }}}
         $\dots$\;
	}
\caption{Attractor computation using summaries.}
\label{alg:use-summaries}
\end{algorithm}

\begin{example}\label{ex:summary-usage}
Recall \Cref{ex:intro-buchi}.
The enforcement summary for the system player described informally there has support location $\mathit{loop}$.
The summary statement is
$\varphi := \exists m_c, m_y. ((c = m_c \land y = m_y) \to \nextp_{\mathit{loop}}(x, y, c)) \land c = m_c$.
If we apply this for attractor $a$ with $a(\mathit{loop}) = (c \geq -199) \land y \leq 0$, we get 
$\exists m_c, m_y. ((c = m_c \land y = m_y) \to (c \geq -199 \land y \leq 0)) \land c = m_c$
which simplifies to $c \geq -199$.
Hence, we set $a(\mathit{loop}) := (c \geq -199)$ and reach a fixpoint in $\mathit{loop}$ after a single step.\looseness=-1
\end{example}

\subsection{Computation of Enforcement Summaries}

Our game-solving method computes enforcement summaries for a given location based on a template characterization of $D$.
It does so by performing attractor computation on the template, keeping the meta-variables constant.
The later is done by \emph{lifting} $\symgame = (L, \linit,\inputs, \progvars, \dom, \delta)$ by meta-variables $\metavars \subseteq \vars$ to the symbolic game structure
$\symgame\uparrow\metavars = (L, \linit, \inputs, \progvars \cup \metavars, \dom, \delta_C)$ with $\delta_C(l, l') := \delta(l,l') \land \bigwedge_{m \in \metavars} m = m'$.
The result of this attractor computation in the aforementioned location is then the summary statement, which is formally stated in the next lemma.

\begin{lemma}\label{lem:summary-computation}
Let $\symgame$ be a symbolic game structure as above, $\metavars \subseteq \vars$ meta-variables, and $\tau: L \to \FOL{\progvars \cup \metavars}$ a template.
Note that $\tau$ is an element of the symbolic states of $\symgame\uparrow\metavars$.
If for $p \in \{\env, \sys\}$ and $l_s \in L$,
    $\sema{\psi} \subseteq \mathit{Attr}_{\symgame\uparrow\metavars,p}(\sema{\tau})(l_s)$
holds with $\psi \in \FOL{\progvars \cup \metavars}$,
then $(p,l_s,\varphi, D)$ with
\\$\varphi := \exists \metavars. \; \psi \land \bigwedge_{l \in L}\forall \progvars.~(\tau (l) \to \nextp_l)$
is an enforcement summary in $\symgame$.
\end{lemma}
When applying $\varphi$ by replacing $\nextp_l$ with some actual symbolic state, it is helpful to apply quantifier elimination.
As $\psi$ can be an under-approximation of the attractor to the template, we can do the summary computation as an anytime computation. That is,  within a given budget, we try to get as good a summary as possible, and we restrict the computation to a part of the overall game.
Furthermore, as the computation reduces to a normal symbolic attractor computation, we can use all available acceleration techniques.
Additionally, this makes summary computation very amenable to being done on demand as needed. 
Next,  we describe how we utilize this approach to compute the templates.

\paragraph{Template Computation.}
For a support location $l_s$ and a current attractor subset $a \in \symstates$,  we compute a template by generalizing $a$.  This helps with obtaining summaries that are more broadly applicable. To this end, we analyze the game around $l_s$ to identify variables  $G \subseteq \progvars$, over which to  generalize.
For a location subset $L_S \subseteq L$, variables  $G \subseteq \progvars$, and current attractor subset $a \in \symstates$ we derive a template as follows.
We introduce one meta-variable per location in $L_S \setminus \{l_s\}$ and variable in $G$, i.e.\ $\metavars := \{ m^l_x \mid l \in L_S\setminus\{l_s\}, x \in G\}$, and define  $\tau$  for all $l \in L$:\looseness=-1
\[ \tau(l) := \textsc{if } l \in L_S\setminus\{l_s\} \textsc { then } \textsc{QELIM}(\exists G. a(l)) \land \left( \bigwedge_{x \in G} (x = m^l_x) \right) \textsc{ else } \bot.
\]
The first conjunct\footnote{$\textsc{QELIM}(\cdot)$  is the result of quantifier elimination applied to the argument.}
states the most general condition for the variables $\progvars \setminus G$, such that we can still apply the template in  $a$.
The second part characterizes (parameterized) assignments for the variables in $G$.  Summaries where individual assignments can be enforced are easy to apply generally.

To select $L_S$, we start with the support location $l_s$ in question and add more locations reachable from it, starting with smaller (local) sets then moving to more global ones.
We select the variables for $G$, according to a novel notion of \emph{point-enforceable variables}.
Intuitively, a variable $x$ is point-enforceable if player~$p$ can, for some states, enforce for $x$ to reach any specific value.
Formally,  $x \in \progvars$ is point-enforceable by player~$p$ in $L_S \subseteq L$, if for all $l \in L_S$ the following is true
\[\exists \progvars \setminus \{x\}. \forall c. \exists x. \cpre{\symgame\uparrow\{c\},p}{\lambda l'.~\textsc{if } l' \in \mathit{Successor}_\symgame(l) \textsc{ then } x = c \textsc{ else } \bot} \]
where $c$ is a fresh variable and $\mathit{Successor}_\symgame(l) \subseteq L$ are the successors of $l$.\looseness=-1

These variables are useful for generalization, as they allow us to propagate equalities in the proposed templates, and with this, generalize over arbitrary conditions on those variables.
This notion is more general than that of  \emph{independent variables} (i.e., variables that do not change) from~\cite{SchmuckHDN24}.

\begin{remark}
Unlike the precomputed attractor acceleration caches in~\cite{SchmuckHDN24},  our enforcement summaries are computed during the actual game-solving.  As a consequence, the result is in general better suited for subsequent applications.
Also, note that enforcement summaries are conceptually more general than attractor acceleration caches and are computed differently.
\end{remark}

\begin{example}\label{ex:summary-computation}
In our example,  for $l_s = \mathit{loop}$ and $a$ with $a(\mathit{iter}) = \top$ we select $L_S = \{\mathit{loop}, \mathit{iter}\}$ and $G = \{c\}$ and obtain a template where $\tau(\mathit{iter}) = (c = m_c^\mathit{iter}) \land (x = m_x^\mathit{iter}) \land (y = m_y^\mathit{iter})$. The computation in the game $\symgame\uparrow\metavars$ results in attractor mapping $\mathit{loop}$ to $c = m_c^{iter}$,  from which we obtain the  summary statement shown in \Cref{ex:summary-usage}.
\end{example}

%% file: experiments.tex
\input{experiments-data}

We implemented\footnote{The implementation, all benchmarks, the data of the results, and the competing tools are available in the artifact at \url{https://doi.org/10.5281/zenodo.18163658}.} our acceleration method in the open-source tool \issy~\cite{HeimD25b}.
\issy\ implements different symbolic solving methods and versions of attractor acceleration and allows writing benchmarks as temporal logic formulas and games.

We compare our method (with and without summaries) against
the standard version of \issy{},
\issy{} without acceleration,
\sweap~\cite{AzzopardiSPS25}, 
\syntheos~\cite{RodriguezGS25}, and
\muval~\cite{UnnoTGK23}. 
The last three solve LTL objectives over program arenas, $\mathit{LTL}^\mathcal{T}$, and fixpoint equations, respectively.
As \cite{HeimD25b}, we do not use \cite{FarzanK18,SamuelDK21,ChoiFPS22,SamuelDK23,RodriguezS23,RodriguezS24,RodriguezGS24} as their prototypes are unavailable, not usable, or are outperformed by the state-of-the-art.
We also do not include the prototypes from~\cite{HeimD24,SchmuckHDN24,HeimD25} as those are integrated in \issy{}.
We also omit~\cite{MaderbacherB22} as it does not have any acceleration techniques and \issy{} outperforms it~\cite{HeimD25b}.

\paragraph{Benchmarks.}%
We use
76 \texttt{rpg} benchmarks from~\cite{NeiderT16,HeimD24,SchmuckHDN24},
56 \texttt{tslmt} benchmarks from~\cite{NeiderT16,MaderbacherB22,HeimD25},
22 \texttt{prog} benchmarks from~\cite{NeiderT16,AzzopardiSPS25},
41  $\mathit{LTL}^\mathcal{T}$ benchmarks from~\cite{RodriguezS23,RodriguezGS25}, and
81 \texttt{issy} benchmarks from~\cite{HeimD25b}.
As the tools use different formalisms and formats, we implemented automatic encodings and used those if a tool does not accept the native benchmark.
This includes translations from $\mathit{LTL}^\mathcal{T}$ to \texttt{issy}, \texttt{prog} to \texttt{issy}, and from each of \texttt{rpg}, \texttt{tslmt}, and \texttt{issy} to each of \texttt{prog}, $\mathit{LTL}^\mathcal{T}$, and \texttt{hes}, where \texttt{hes} is the format for \muval{}.
We discarded manually translated benchmarks used in other works\footnote{This is why some results in the literature might look fairly different.} as we found that some of those contain serious mistakes, e.g., introducing falsity in the assumptions. %

However, to be semantic preserving, the automatic encoding is sometimes fairly elaborate.
In particular, since \sweap's formalism does not support unbounded inputs or outputs (such as integers), those have to be modeled as multi-step inputs or outputs, resulting in overhead that can affect performance.
It was sometimes unclear what some of the tools accept, with errors occurring late at runtime.
Also, some benchmarks from~\cite{AzzopardiSPS25} are not well-defined according to their semantics, and we choose an interpretation that seems to be different from \sweap{}.
Nevertheless, an automatic encoding is the better scientific choice.

We created new benchmarks for which we expected that more complicated acceleration arguments would be needed than in the existing ones.
Some are abstract and conceptual, while others are application-inspired, and they are both temporal formulas and games. 
We also adapted verification examples from~\cite{LeikeH15}.

\paragraph{Results.}
We ran all experiments on an AMD EPYC processor with 1 core, 6GB of memory, and a 20-minute wall-clock time-out per run.
\Cref{fig:compare-lit} and \Cref{tab:compare-lit-table} show the results on the literature benchmarks. 
We did not use enforcement summaries there, as they would introduce additional overhead. 
However, they are not required for those benchmarks that are already handled well by existing techniques.
\Cref{tab:new-benchmarks} shows the results on the new benchmarks.
If not stated otherwise, we ran the tools (if possible) to check realizability.
We do not distinguish between runtime errors and out-of-memory errors for technical reasons.

\macrodataliteratur

\macrotablefull

\paragraph{Discussion.}
Overall,  the results indicate that our new technique for composing acceleration arguments broadens the range of solvable problems while remaining competitive with the state of the art. 
\Cref{tab:compare-lit-table} shows that from the literature benchmarks, it has the most unique solves. In \Cref{fig:compare-lit} we see that the overall performance on these benchmarks is comparable to the state-of-the-art tool \issy{}. 
The results on new benchmarks reported in \Cref{tab:new-benchmarks} show that our tool can discover complex acceleration arguments and solve games that other tools fail to handle.
The closest competitor, \issy{}, can in some cases find simple arguments where the GAL search explores unfruitful parts of the lemma space.
As expected, the use of summaries provides notable performance benefits on benchmarks where similar GALs are otherwise computed over and over again, such as most of the B\"uchi game benchmarks.
However, summary computation can introduce significant overhead, as our implementation allocates a lot of time for computing summaries and blocks the remaining computation.
This could be mitigated by interleaving summary computation with the main analysis.

In the future, we plan to address the observed performance drawbacks in the current GAL computation.
We aim to develop better strategies for guiding the GAL search, for example, by applying data-flow analysis to the symbolic game.
Other scalability challenges (e.g., for synthesis) stem from the complexity of generated terms and the current limitations of \issy{}'s term simplification mechanism.
We believe incorporating abstraction techniques can be fruitful there.

%% file: experiments-data.tex
\newcommand{\macrocompare}{
\begin{tikzpicture}
\begin{axis}[width=0.9\textwidth, height=3.7cm, ylabel near ticks, legend pos=outer north east, xmin=-5]
\addplot[., red] table[y=Solved,x=Time,col sep=comma] {data/plot-issy-edge.csv};
\addplot[., blue] table[y=Solved,x=Time,col sep=comma] {data/plot-issy-cav25ae.csv};
\addplot[., green] table[y=Solved,x=Time,col sep=comma] {data/plot-issy-no-accel.csv};
\addplot[., brown] table[y=Solved,x=Time,col sep=comma] {data/plot-muval-1d4999.csv};
\addplot[., yellow] table[y=Solved,x=Time,col sep=comma] {data/plot-sweap-artifact.csv};
\addplot[., gray] table[y=Solved,x=Time,col sep=comma] {data/plot-syntheos-c8ef82e.csv};
\addlegendentry{\scriptsize New tech.}
\addlegendentry{\scriptsize \issy}
\addlegendentry{\scriptsize \issy\ no acc.}
\addlegendentry{\scriptsize \muval}
\addlegendentry{\scriptsize \sweap}
\addlegendentry{\scriptsize \syntheos}
\end{axis}
\end{tikzpicture}}
\newcommand{\macrotablesum}{
\begin{tabular}{|l|r|rr|rr|} 
\hline
Name                  & ~Tot~ & Sol   & Un    & TO    & NR    \\\hline
New tech.             & 239   & 195   & 13    & 40    & 4     \\
\issy\                & 239   & 191   & 3     & 32    & 16    \\
\issy\ no acc.        & 239   & 113   & 1     & 71    & 55    \\
\sweap\               & 206   & 20    & 6     & 59    & 127   \\
\muval\               & 239   & 84    & 4     & 80    & 75    \\
\syntheos\            & 180   & 47    & 0     & 37    & 96    \\\hline
\end{tabular}}

\newcommand{\macrodataliteratur}{%
\begin{figure}[t!]
\begin{minipage}[t!]{0.47\textwidth}
\centering
\scalebox{0.8}{\macrocompare{}}
\captionof{figure}{\footnotesize Solved literature benchmarks (out of 239) within time in seconds. The new technique is without summaries.\\}\label{fig:compare-lit}
\end{minipage}
\hfill
\begin{minipage}[t!]{0.47\textwidth}
\centering
\scalebox{0.8}{\macrotablesum{}}
\captionof{table}{\footnotesize On 239 literature benchmarks, \textbf{Tot}al applicable, \textbf{Sol}ved, \textbf{Un}iquely solved (full list in \Cref{app:additional-data}), \textbf{T}ime\textbf{O}ut, \textbf{N}o\textbf{R}esult (MO/err.)}\label{tab:compare-lit-table}
\end{minipage}
\end{figure}}

\newcommand{\macrotablefull}{
\begin{table}[t!]
\scriptsize
\begin{minipage}[t]{0.48\textwidth}
\centering
\scalebox{0.73}{%
\begin{tabular}{|ll||rr|rrrrr||rr|} 
\hline
Name                           & W     & NT     & SU     & I      & A      & W      & Y      & M      & NT-S  & SU-S  \\\hline
buchi                          & R     &\be{63} & T      & T      & T      & T      & N      & T      & 95    & T     \\
buchi-simple                   & R     &\be{19} & T      & T      & T      & N      & T      & T      & 38    & T     \\
choice-3-actions               & R     &\be{1}  & 2      & 4      & T      & T      & N      & T      & 213   & N     \\
choice-4-actions               & R     &\be{1}  & 2      & 9      & T      & T      & N      & T      & N     & T     \\
choice-actions-i               & R     &\be{16} & T      & T      & T      & T      & N      & T      & 599   & T     \\
equal-mod2-real                & R     &\be{1}  & T      & 2      & T      & N      & T      & T      & 4     & T     \\
fault-tolerance                & R     &\be{4}  & T      & T      & T      & N      & -      & T      & T     & T     \\
lemma-chain-control            & R     &\be{1}  & 2      & T      & T      & N      & -      & T      & T     & N     \\
lemma-chain                    & R     &\be{1}  & 2      & T      & T      & N      & -      & T      & 73    & T     \\
nested-x-y-z                   & R     &\be{39} & 42     & T      & T      & N      & N      & T      & T     & T     \\
nondet-exit-swap-in            & R     &\be{1}  & 2      & 3      & T      & T      & -      & T      & 485   & N     \\
nondet-exit-swap               & R     &\be{1}  & 2      & 3      & T      & T      & -      & T      & 5     & T     \\
prevent-zeno                   & R     &\be{3}  & 12     & N      & T      & -      & N      & N      & N     & T     \\
ranking-choice-2               & R     &\be{4}  & T      & T      & T      & T      & N      & T      & 78    & T     \\
ranking-choice-3               & R     &\be{26} & T      & T      & T      & T      & N      & T      & 327   & T     \\
ranking-choice-4               & R     &\be{507}& T      & T      & T      & T      & N      & T      & T     & T     \\
transfer-lin-constr.           & R     &\be{1}  & T      & T      & T      & -      & N      & N      & T     & T     \\
uav-chaotic-unreal             & U     &\be{5}  & 7      & T      & T      & -      & N      & N      & -     & -     \\\hline
bu-hard-loop-200               & U     & 282    &     62 & T      & T      & N      & -      &\be{24} & -     & -     \\
bu-hard-loop-400               & U     & 784    &    121 & T      & T      & N      & -      &\be{14} & -     & -     \\
bu-hard-loop-800               & U     & T      &    241 & T      & T      & N      & -      &\be{29} & -     & -     \\
equal-mod2-unreal              & U     & T      & T      & T      & T      & N      & N      &\be{39} & -     & -     \\
inequality-assump.             & R     & 15     & T      & \be{5} & T      & T      & N      & T      & 41    & T     \\
prio-tasks-real-100            & R     & T      & 1125   &\be{378}& T      & T      & N      & T      & T     & T     \\
prio-tasks-real-200            & R     & T      & T      &\be{758}& T      & T      & N      & T      & T     & T     \\
prio-tasks-unre-100            & U     & 58     & 8      & \be{2} & T      & T      & N      & 288    & -     & -     \\
rect-patrol-buchi              & R     & T      & T      & N      &\be{29} & -      & N      & N      & T     & T     \\
service-2                      & R     & 14     & T      & \be{11}& T      & N      & N      & T      & 73    & T     \\
service-4                      & R     & 77     & T      & \be{24}& T      & N      & N      & N      & 480   & T     \\\hline
\end{tabular}}
\end{minipage}
\hfill
\begin{minipage}[t]%
{0.48\textwidth}
\centering
\scalebox{0.73}{%
\begin{tabular}{|ll||rr|rrrrr||rr|} 
\hline
Name                           & W     & NT     & SU     & I      & A      & W      & Y      & M      & NT-S  & SU-S  \\\hline
service-8                      & R     & 1160   & T      & \be{45}& T      & N      & N      & N      & T     & T     \\
service-bounds-10              & R     & 167    & T      & \be{19}& T      & N      & N      & T      & 609   & T     \\
service-bounds-20              & R     & 166    & T      & \be{19}& T      & N      & N      & T      & 608   & T     \\\hline
double                         & R     & T      & T      & T      & T      & N      & N      & T      & T     & T     \\
equality-assump.               & R     & T      & T      & T      & T      & T      & N      & T      & T     & T     \\
nested-x-y-z-u                 & R     & T      & T      & T      & T      & N      & N      & T      & T     & T     \\
nested-x-y-z-u-v               & R     & T      & T      & T      & T      & N      & T      & T      & T     & T     \\
service-10                     & R     & T      & T      & T      & T      & N      & N      & N      & T     & T     \\\hline
bu-loop-200                    & R     & 96     &\be{55} & T      & T      & N      & -      & T      & 111   & 68    \\
bu-loop-400                    & R     & 190    &\be{108}& T      & T      & N      & -      & T      & 218   & 126   \\
bu-loop-800                    & R     & 381    &\be{216}& T      & T      & N      & -      & T      & 430   & 248   \\
bu-loop-chain-200              & R     & 813    &\be{55} & T      & T      & T      & -      & T      & 1133  & N     \\
bu-loop-chain-400              & R     & T      &\be{108}& T      & T      & N      & -      & T      & T     & N     \\
bu-loop-chain-800              & R     & T      &\be{215}& T      & T      & N      & -      & T      & T     & N     \\
bu-loop-swap-200               & R     & 512    &\be{55} & 709    & T      & T      & -      & T      & T     & N     \\
bu-loop-swap-400               & R     & 1014   &\be{111}& T      & T      & T      & -      & T      & T     & N     \\
bu-loop-swap-800               & R     & T      &\be{220}& T      & T      & T      & -      & T      & T     & N     \\
reach-either-or                & R     &\be{1}  &\be{1}  & T      & T      & N      & -      & T      & 2     & N     \\
torus-game                     & R     & 21     & \be{19}& 92     & T      & -      & N      & N      & N     & T     \\
uav-chaotic                    & R     & T      & \be{38}& T      & T      & -      & N      & N      & T     & N     \\\hline
ex-2-05~(\textit{adapted}      & R     &\be{1}  & 2      & 2      & T      & N      & T      & T      & 2     & N     \\
ex-4-06~~\textit{from}         & R     &\be{1}  & 2      &\be{1}  & T      & N      & T      & T      & 6     & N     \\
ex-4-11~~\cite{LeikeH15})      & R     &\be{1}  & T      & 4      & T      & N      & T      & T      & T     & T     \\
ex-4-12                        & U     & T      & T      & T      & T      & T      & T      &\be{98} & -     & -     \\
ex-4-13                        & R     & T      & T      & T      & T      & T      & N      & T      & T     & T     \\
ex-4-15                        & R     & T      & T      & T      & T      & T      & N      & T      & T     & T     \\
ex-4-18                        & R     &\be{1}  & 2      & T      & T      & N      & -      & T      & 353   & 356   \\
ex-4-21                        & R     &\be{1}  & 32     & 1      & T      & N      & -      & T      & 4     & T     \\
ex-5-01                        & R     &\be{110}& 115    & T      & T      & N      & N      & T      & T     & N     \\\hline
\end{tabular}}
\end{minipage}
\vspace{3mm}
\caption{\footnotesize Run on new benchmarks which are \textbf{R}ealizable or \textbf{U}nrealizable of the \textbf{N}ew \textbf{T}echnique, new technique with \textbf{SU}mmaries, \textbf{I}\texttt{ssy}, \issy\ without \textbf{A}cceleration,  \texttt{s}\textbf{W}\texttt{eap}, \texttt{S}\textbf{Y}\texttt{ntheos}, \textbf{M}\texttt{uVal}, and synthesis variants (*\textbf{-S}), depicting resulting runtime in seconds, \textbf{T}imeout, \textbf{N}o result, or not applicable (\textbf{-}). The runs were grouped retrospectively.}\label{tab:new-benchmarks}
\vspace{-5mm}
\end{table}}

%% file: appendix-proofs.tex
\newcommand{\mcombvvent}{\combine{\assignment}{\assignment'} \FOLentailsT{T}}
\newcommand{\mvent}{\assignment \FOLentailsT{T}}
\newcommand{\mvpent}{\assignment' \FOLentailsT{T}}
\newcommand{\malent}[1]{\combine{\alpha[#1]}{\alpha[#1+1]} \FOLentailsT{T}}

\subsection{Proofs of \Cref{sec:lemmas}}

\begin{proof}[\Cref{lem:gal-intersect}]
Let $L_0 = (\base_0, \stay_0, \step_0, \conc_0)$ and\\
 $L_1 = (\base_1, \stay_1, \step_1, \conc_1)$ be GALs over the same variables $V$. 
Furthermore, let $\base := \base_0 \land \base_1$, $\stay := \stay_0 \land \stay_1 \land{stayBase}$, $\conc := \conc_0 \land \conc_1$,
$\step := \mathit{stayBase} \land \bigvee_{i \in \{0, 1\}}(\step_i \land \lnot\base_i \land \stay_{1 - i})$, and
$\mathit{stayBase} := \bigwedge_{i \in \{0,1\}}(\base_i \land \lnot \base_{1 - i} \to \base_i[\progvars \mapsto \progvars'])$.
To prove that\\$L = (\base, \stay, \step, \conc)$ is a GAL we check the conditions in \Cref{def:gen-lemma}.

\paragraph{Condition (II)}
Let $\assignment, \assignment' \in \assignments{V}$ be assignments such that $\mcombvvent \step \vee \stay$.
By definition of $\stay$ and $\step$, either $\mcombvvent \stay_0 \land \stay_1$, or $\mcombvvent \step_0 \land \stay_1$, or $\mcombvvent \step_1 \land \stay_0$ holds.
Hence for $n \in \{0,1\}$, $\mcombvvent \step_n \lor \stay_n$ holds.
As by assumption that $L_n$ is a GAL, this implies that if $\mvent \conc_n$, then $\mvpent \conc_n$ holds.
Hence, if $\mvent \conc_0 \land \conc_1$, then $\mvpent \conc_0 \land \conc_1$ which establishes (II) for $L$.

\paragraph{Condition (I)}
Let $\alpha \in \assignments{V}^\omega$ such that 
\begin{enumerate}
    \item[(z)]   $\alpha[0] \FOLentailsT{T} \conc$, and
    \item[(a)]   for all $i \in \Nat$, $\malent{i} \step \lor \stay$ holds, and
    \item[(b)]   for all $i \in \Nat$ there exists $j \geq i$ such that $\malent{j} \step$.
\end{enumerate}
We now need to prove that there exists $k \in \Nat$ such that, $\alpha[k] \FOLentailsT{T} \base_0 \land \base_1$.

Note that (b) implies that  $\malent{i} \mathit{stayBase}$ holds for all $i \in \Nat$.
Hence, to prove our goal, it suffices that prove that there exists $k_0, k_1 \in \Nat$, such that  $\alpha[k_n] \FOLentailsT{T} \base_n$ for both $n \in \{1, 2\}$.

Now let $n \in \{0, 1\}$.
As $L_n$ is a GAL we try to prove the former by applying condition (I) of $L_n$ to $\alpha$.
First note that the definition of $\conc$ and (z) imply that $\alpha[0] \FOLentailsT{T} \conc_n$.
Now let $i \in \Nat$. 
As for condition (II), the definitions of $\stay$ and $\step$ imply with (a) that also $\malent{i} \stay_n \lor \step_n$ holds.
It remains to prove that for all $i \in \Nat$ if there exists $j_n \geq i$ such that $\malent{j} \step_n$ holds.
Assume towards contradiction that for all $j_n \geq i$, $\malent{j} \step_n$ does not hold. 
By (b) there exists $j \geq j_n$ such that $\malent{j} \step$ holds.
As by assumption $\malent{j} \step_n$ also does not hold (we assumed towards contradiction \emph {for all} $j_n \geq i$), we conclude by the definition of $\step$ that $\malent{j} \step_{1 - n} \land \lnot \base_{1 - n}$ has to hold.
However, this now implies that for all $i \in \Nat$, there exists $j \geq j_n \geq i$, such that 
$\malent{j} \step_{1 - n}$ and $\malent{j} \lnot \base_{1 - n}$ hold.
With our previous observations, the first implies all conditions of (I) in $L_{1 -n}$ apply to $\alpha$ such that $k_{1 - n}$ exists such that $\alpha[k_{n - 1}] \FOLentailsT{T} \base_{1 - n}$.
With our observation on $\mathit{stayBase}$, this contradicts the second part. 
\qed
\end{proof}

\begin{proof}[\Cref{lem:gal-union}]
Let $L_0 = (\base_0, \stay_0, \step_0, \conc_0)$ and\\
$L_1 = (\base_1, \stay_1, \step_1, \conc_1)$ be GALs over the same variables $V$. 
Furthermore, let $\base := \base_0 \lor \base_1$, $\stay := \stay_0 \land \stay_1$, $\conc := \conc_0 \lor \conc_1$, and 
$\step := (\conc_0 \land \step_0 ) \lor (\conc_1 \land \step_1 \land \stay_0)$.
To prove that $L = (\base, \stay, \step, \conc)$ is a GAL we check the condition in \Cref{def:gen-lemma}.

\paragraph{Condition (II)}
Let $\assignment, \assignment' \in \assignments{V}$ be assignments such that 
$\mcombvvent \step \vee \stay$.
By definition of $\step$ and $\stay$ the second implies that
$\mcombvvent \conc_0 \land \step_0$, or
$\mcombvvent \conc_1 \land \step_1 \land \stay_0$, or
$\mcombvvent \stay_0 \land \stay_1$, holds.
Now let $\mvent \conc$ also hold. 
If $\mvent \conc_0$ then condition (II) follows in all three cases by that $L_0$ is a GAL.
If $\mvent \conc_1$ then condition (II) follows in the second and third case by that $L_1$ is a GAL.
The first case implies $\mvent \conc_0$ which we already cover.

\paragraph{Condition (I)}
Let $\alpha \in \assignments{V}^\omega$ such that $\alpha[0] \FOLentailsT{T} \conc$, and
\begin{enumerate}
    \item[(a)]   for all $i \in \Nat$, $\malent{i} \step \lor \stay$ holds, and
    \item[(b)]   for all $i \in \Nat$ there exists $j \geq i$ such that $\malent{j} \step$.
\end{enumerate}
We now need to prove that there exists $k \in \Nat$ such that, $\alpha[k] \FOLentailsT{T} \base_0 \lor \base_1$.
The definition of $\step$ and (b) create two cases: 
\begin{enumerate}
    \item[(A)]    Either $\malent{j} \conc_0 \land \step_0$ holds infinitely often, i.e. for all $i \in \Nat$ there exists $j \geq i$ such that $\malent{j} \conc_0 \land \step_0$ holds. 
    \item[(B)]    Form some point on it stops holding, i.e.,  there exists $C \in \Nat$, such that for all $i \geq C$, there exists $j \geq i$ such that $\malent{j} \conc_1 \land \step_1$ and with (a) $\malent{i} \stay_1 \lor \step_1$. 
\end{enumerate}

In case (A) follows that there exists $C \in \Nat$ such that $\alpha[C] \FOLentails \conc_0$. 
Furthermore, the definitions of $\step$ and $\stay$ imply with (a) that for all $i \in \Nat$, $\malent{i} \step_0 \lor \stay_0$ stay holds.
As $L_0$ is a GAL, the condition (I) there in $L_0$ applies now to the shifted sequence $\lambda k.\alpha[k + C]$.
Hence, there exist $k \in \Nat$ such that $\alpha[k + C] \FOLentailsT{T} \base_0$ which proves our claim.

In case (B) we can assume w.l.o.g.\ that $\malent{C} \conc_1 \land \step_1$ holds, as it does eventually by the first part in (B).
As this implies $\malent{C} \conc_1$ we cleared all conditions to apply (I) of GAL $L_1$ onto $\lambda k.\alpha[k + C]$ and exists $k \in \Nat$ such that $\alpha[k + C] \FOLentailsT{T} \base_1$.
\qed
\end{proof}

\begin{proof}[\Cref{lem:gal-chain}]
Let $L_0 = (\base_0, \stay_0, \step_0, \conc_0)$ and\\ $L_1 = (\base_1, \stay_1, \step_1, \conc_1)$ be GALs over the same variables $V$. 
Let
$\stay :=  \stay_0 \land \stay_1$ and 
$\stay := \stay_0 \land \stay_1 \land (\base_1 \to \base_1[\progvars \mapsto \progvars'])$
To prove that $L = (\base_0, \stay, \step, \conc)_0$ is a GAL we check the conditions in \Cref{def:gen-lemma}.

\paragraph{Condition (II)}
Let $\assignment, \assignment' \in \assignments{V}$ be assignments such that 
$\mvent \conc_0$ and
$\mcombvvent \step \vee \stay$.
By definition of $\step$ and $\stay$ the second implies that
$\mcombvvent \step_0$ or $\mcombvvent \stay_0$.
As $L_0$ is a GAL, this implies that $\mvpent \conc_0$ which establishes this condition.

\paragraph{Condition (I)}
Let $\alpha \in \assignments{V}^\omega$ such that $\alpha[0] \FOLentailsT{T} \conc_0$, and
\begin{enumerate}
    \item[(a)]   for all $i \in \Nat$, $\malent{i} \step \lor \stay$ holds, and
    \item[(b)]   for all $i \in \Nat$ there exists $j \geq i$ such that $\malent{j} \step$.
\end{enumerate}
We now need to prove that there exists $k \in \Nat$ such that, $\alpha[k] \FOLentailsT{T} \base_0$.
As shown for the previous condition (a) and the definitions $\step$ and $\stay$ imply that for all $i \in \Nat$, $\malent{i} \step_0 \lor \stay_0$ holds.
As $L_0$ is a GAL we can use (I) there the prove our goal. 
Hence, it remains to show that for all $i \in \Nat$ there exists $j \geq i$ such that $\malent{j} \step_0$. 

Assume towards contradiction, that there exists $C \in \Nat$ such that for all $j \geq i$, $\malent{j} \step_0$ does not hold.
In combination with (b) and the definition of $\step$ this implies that for all $i \geq C$ there exists $j \geq i$ such that $\malent{j} \step_1 \land \conc_1 \land \base_1$.
Let $D \geq C$ be the first point where this is that case, then $\alpha[D] \FOLentailsT{T} \conc_1$. 
Furthermore, for all $i \geq D$ there exists $j \geq i$ such that $\malent{j} \step_1$ holds and by (a), the definition of $\stay$ and the assumption, also for all $i \geq D$, $\malent{i} \step_1 \lor \stay_1$ holds.
Hence, by applying condition (I) of the GAL $L_1$ to the shifted sequence $\lambda k.\alpha[k + D]$, this implies there exists $k \in \Nat$ such that $\alpha[k + D] \FOLentailsT{T} \base_1$. 
Note that if for some $\alpha[k + D + m] \FOLentailsT{T} \base$ holds for $m \in \Nat$, then $\step$ is not applicable anymore, i.e., $\malent{k + D + m} \step$ does not hold, which implies that $\malent{k + D + m} \step$ holds, which in turn implies by the Defintion of $\step$, that $\alpha[k + D + m + 1] \FOLentailsT{T} \base$ holds. 
Hence, by induction, $\alpha[k + D + m] \FOLentailsT{T} \base$ holds for all $m \in \Nat$, as it holds for $m = 0$.
This implies that for all $m \in \Nat$, $\malent{k + D + m} \step$ does not hold, which implies that  $\malent{k + D + m} \step_1$ does not hold, which poses a contradiction. 
\qed
\end{proof}

\begin{proof}[\Cref{lem:gal-inv}]
Let $L = (\base, \stay, \step, \conc)$ be a GAL over $V$, $\inv \in \FOL{V}$ be an invariant, and $\inv' := \inv[V \mapsto V']$.
To prove that 
$L_I = (\base \land \inv, \stay \land \inv', \step \land \inv', \conc \land \inv)$ is a GAL we check the conditions in \Cref{def:gen-lemma}.
For condition (I), let $\assignment, \assignment' \in \assignments{V}$ be assignments such that 
$\mvent \conc \land \inv$ and
$\mcombvvent \stay \land \inv' \lor \step \land \inv'$.
This implies directly that $\mvpent \inv$, and since $L$ is a GAL also $\mvpent \conc$, which concludes this condition.
For condition (II), let $\alpha \in \assignments{V}^\omega$ such that
\begin{enumerate}
    \item[(z)] $\alpha[0] \FOLentailsT{T} \conc \land \inv$ holds, and
    \item[(a)] for all $i \in \Nat$, $\malent{i} \stay \land \inv' \lor \step \land \inv'$ holds, and
    \item[(b)] for all $i \in \Nat$ there exists $j \geq i$ such that $\malent{j} \step \land \inv'$.
\end{enumerate}
We now need to prove that there exists $k \in \Nat$ such that, $\alpha[k] \FOLentailsT{T} \base \land \inv$.
By natural induction it follows with conditions (z) and (a) that for all $i \in \Nat$, $\alpha[i] \FOLentailsT{T} \land \inv$ holds.
Furthermore, $\alpha$ fulfills all preconditions of (I) in the GAL $L$ such that there exists $k \in \Nat$ such that $\alpha[k] \FOLentailsT{T} \base$.
Hence $\alpha[k] \FOLentailsT{T} \base \land \inv$ also holds which proves the claim.
\qed
\end{proof}

\begin{proof}[\Cref{lem:gal-base}]
Let $t$ be a term over numeric variables $V$ and
let $a \in \Real \cup \{-\infty\}$, $b \in \Real \cup \{\infty\}$ with $a \leq b$ be bounds.
Furthermore, let $t' := t[V \mapsto V']$, $\epsilon > 0$, and
\begin{itemize}
    \item   $\base := a \leq t \leq b$
    \item   $\conc := \top$,
    \item   $\stay := (a \leq t' \leq b) \lor (t < l \land t \leq t' \leq b) \lor (t > a \land t \geq t' \geq a)$, and
    \item   $\step := (a \leq t' \leq b) \lor (t < l \land t + \epsilon \leq t' \leq b) \lor (t > a \land t - \epsilon \geq t' \geq a).$
\end{itemize}
To prove that $(\base, \stay, \step, \conc)$ is a GAL we check \Cref{def:gen-lemma}.

Condition (II) is trivial.
For (I), let $\alpha \in \assignments{V}^\omega$ such that
\begin{enumerate}
    \item[(a)] for all $i \in \Nat$, $\malent{i} \stay \lor \step$ holds, and
    \item[(b)] for all $i \in \Nat$ there exists $j \geq i$ such that $\malent{j} \step$.
\end{enumerate}
We now need to prove that there exists $k \in \Nat$ such that, $\alpha[k] \FOLentailsT{T} a \leq t \leq b$.
Note that if $a = \infty$ and $b = -\infty$ this is trivial, as well as if, $\alpha[0] \FOLentailsT{T} a \leq t \leq b$
We consider the case where $\alpha[k] \FOLentailsT{T} t > b$, the case $\alpha[k] \FOLentailsT{T} t < a$ works analogously.
First observe that (a) implies inductively, that for all $i \in \Nat$, if $\alpha[i] \not\FOLentails{T} a \leq t \leq b$, then $\alpha[i] > b$.
Hence, as long as $\alpha[i]$ does not reach $\base$, it holds that $\stay$ implies $t \geq t'$ and $\step$ implies that $t - \epsilon \geq t'$.
Now in this case conditions (a) and (b) imply that the evaluation of term $t$ on $\alpha$, do not increase and infinitely often decrease by a fixed $\epsilon$ which means that eventually $t \leq b$ will be reached which proves the claim. 
\qed
\end{proof}

\subsection{Proofs of \Cref{sec:summaries}}

\begin{proof}[\Cref{lem:summary-computation}]
Let $\symgame = (L, \linit,\inputs, \progvars, \dom, \delta)$ be a symbolic game structure, $\metavars \subseteq \vars$ meta-variables, and $\tau: L \to \FOL{\progvars \cup \metavars}$ a template.
Let $\symgame_C = (L, \linit, \inputs, \progvars \cup \metavars, \dom, \delta_C)$ with $\delta_C(l, l') := \delta(l,l') \land \bigwedge_{m \in \metavars} m = m'$ be a modification of $\symgame$ which includes $\metavars$ as constant variables. 

Let $p \in \{\env, \sys\}$ and $l_s \in L$ and consider 
$\psi \in \FOL{\progvars \cup \metavars}$ such that 

\[\sema{\psi} \subseteq \mathit{Attr}_{\symgame_C,p}(\sema{\tau})(l_s).\]

Furthermore,  let 

\[\varphi := \exists \metavars. \; \psi \land \bigwedge_{l \in L}\forall \progvars.~(\tau (l) \to \nextp_l(\progvars)).\]

We have to show that 
$(p,l_s,\varphi, D)$ 
is an enforcement summary in $\symgame$, that is, it satisfies the condition of \Cref{def:summary}.  
Let $d \in D$ and let $\assignment \in \sema{ \varphi\llparenthesis\nextp_l \mapsto d(l) \mid l \in L\rrparenthesis}$.  We have to show that $\assignment \in \mathit{Attr}_{\symgame,p}(\sema{d})(l_s)$.

By the definition of $\varphi$ we have that  
$\assignment  \FOLentailsT{T}   \exists \metavars. \; \psi \land \bigwedge_{l \in L}\forall \progvars.~(\tau (l) \to d(l))$.
Thus, there exists $\assignment_{\metavars}\in \assignments{\metavars}$ such that 
\begin{itemize}
\item[(1)] $\assignment \uplus \assignment_{\metavars}  \FOLentailsT{T}  \psi$, and
\item[(2)] $\assignment \uplus \assignment_{\metavars}  \FOLentailsT{T}  \bigwedge_{l \in L}\forall \progvars.~(\tau (l) \to d(l))$, i.e., $\assignment_{\metavars}  \FOLentailsT{T}  \bigwedge_{l \in L}\forall \progvars.~(\tau (l) \to d(l))$. 
\end{itemize}
 Therefore, by the choice of $\psi$, we have that 
\[\assignment \uplus \assignment_{\metavars} \in \mathit{Attr}_{\symgame_C,p}(\sema{\tau})(l_s).\]

By the definition of attractor,  this implies that 
there exists $\sigma \in \str{p}{\sema{\symgame_C}}$ such that  for all
$\pi \in \plays_{\sema{\symgame_C}}((l_s,\assignment \uplus \assignment_{\metavars}),\sigma)$ there  exists $n$ such that $\pi_n\in \sema{\tau}$.
Since the meta-variables in $\metavars$ are constant in $\symgame_C$, we have that for every such $\pi$ and $n$,  if $\pi_n = (l',\assignment' \uplus \assignment'_{\metavars})$, then $\assignment'_\metavars = \assignment_{\metavars}$. 
By (2) above, we have that 
$\assignment'_{\metavars}  \FOLentailsT{T}   \bigwedge_{l \in L}\forall \progvars.~(\tau (l) \to d(l))$. 
Since $\pi_n = (l',\assignment' \uplus \assignment'_{\metavars})\in \sema{\tau}$, this means that $\assignment'  \FOLentailsT{T}  d(l')$.
This implies that the strategy $\sigma$ for player $p$ in $\sema{\symgame_C}$  enforces from $(l_s,\assignment \uplus \assignment_{\metavars})$ reaching the set of states $\{(l',\assignment'\uplus \assignment_{\metavars})\mid (l',\assignment') \in \sema{d}\}$.

By the construction of $\symgame_C$ from $\symgame$, we have that any strategy in $\symgame_C$ can be transformed to a strategy in $\symgame$ by ignoring the values of the meta-variables. 
The plays of the resulting strategy in $\symgame$
are obtained from the plays of the strategy in  $\symgame_C$ by ignoring the values of the meta-variables,  and thus they visit the exact same sequences of states in $\symgame$.  Therefore,  what we showed above implies that there 
exists a  strategy $\sigma{_\symgame}$ for player $p$ in $\sema{\symgame}$ that  enforces from $(l_s,\assignment)$ reaching the set $ \sema{d}$.  
Thus,  $(l_s,\assignment) \in  \mathit{Attr}_{\symgame_C,p}(\sema{\tau})$, which is what we had to show.
\qed
\end{proof}

%% file: appendix-more-data.tex
Our new technique solved
\begin{itemize}
\item cav24-schmuck-smarthome-day-cold,
\item cav24-schmuck-smarthome-day-empty,
\item cav24-schmuck-smarthome-day-not-empty,
\item cav24-schmuck-smarthome-day-warm-or-cold,
\item cav24-schmuck-smarthome-day-warm,
\item cav24-schmuck-smarthome-night-empty,
\item cav24-schmuck-smarthome-nightmode,
\item cav24-schmuck-smarthome-night-sleeping,
\item popl25-heim-limitations-precise-reachability,
\item popl25-heim-robot-to-target,
\item popl25-heim-tasks,
\item cav25-heim-test-05, and
\item cav25-heim-two-vars-real
\end{itemize}
uniquely.
\issy{} solved 
\begin{itemize}
\item fmcad22-maderbacher-sort5,
\item popl24-heim-robot-cat-real-1d, and
\item popl25-heim-robot-to-target-charging 
\end{itemize}
uniquely.
\issy{} without acceleration solved
\begin{itemize}
\item fmcad22-maderbacher-elevator-signal-5
\end{itemize}
uniquely.
\sweap{} solved
\begin{itemize}
\item cav25-azzopardi-elevator-paper,
\item cav25-azzopardi-reversible-lane-r,
\item cav25-azzopardi-robot-collect-samples-v4,
\item cav25-azzopardi-robot-grid-reach-repeated-with-obstacles-1d,
\item cav25-azzopardi-robot-grid-reach-repeated-with-obstacles-2d, and
\item cav25-azzopardi-taxi-service,
\end{itemize}
uniquely.
\muval{} solved 
\begin{itemize}
\item popl24-heim-robot-cat-unreal-1d,
\item popl24-heim-robot-cat-unreal-2d,
\item popl25-heim-g-real, and
\item popl25-heim-limitations-buffer-storage,
\end{itemize}
uniquely.

%% file: bib/own-publications.bib
@article{HeimD24,
  author       = {Philippe Heim and
                  Rayna Dimitrova},
  title        = {Solving Infinite-State Games via Acceleration},
  journal      = {Proc. {ACM} Program. Lang.},
  publisher    = {{ACM}},
  volume       = {8},
  number       = {{POPL}},
  pages        = {1696--1726},
  year         = {2024},
  doi          = {10.1145/3632899},
}

@inproceedings{SchmuckHDN24,
  author       = {Anne{-}Kathrin Schmuck and
                  Philippe Heim and
                  Rayna Dimitrova and
                  Satya Prakash Nayak},
  editor       = {Arie Gurfinkel and
                  Vijay Ganesh},
  title        = {Localized Attractor Computations for Infinite-State Games},
  booktitle    = {Computer Aided Verification - 36th International Conference, {CAV} 2024},
  series       = {LNCS},
  volume       = {14683},
  pages        = {135--158},
  publisher    = {Springer},
  year         = {2024},
  doi          = {10.1007/978-3-031-65633-0_7},
}

@article{HeimD25,
  author       = {Philippe Heim and
                  Rayna Dimitrova},
  title        = {Translation of Temporal Logic for Efficient Infinite-State Reactive
                  Synthesis},
  journal      = {Proc. {ACM} Program. Lang.},
  publisher    = {{ACM}},
  volume       = {9},
  number       = {{POPL}},
  pages        = {1536--1567},
  year         = {2025},
  doi          = {10.1145/3704888},
}

@inproceedings{HeimD25b,
  author       = {Philippe Heim and
                  Rayna Dimitrova},
  editor       = {Ruzica Piskac and
                  Zvonimir Rakamaric},
  title        = {Issy: {A} Comprehensive Tool for Specification and Synthesis of Infinite-State
                  Reactive Systems},
  booktitle    = {Computer Aided Verification - 37th International Conference, {CAV}
                  2025, Zagreb, Croatia, July 23-25, 2025, Proceedings, Part {IV}},
  series       = {Lecture Notes in Computer Science},
  volume       = {15934},
  pages        = {298--312},
  publisher    = {Springer},
  year         = {2025},
  url          = {https://doi.org/10.1007/978-3-031-98685-7\_14},
  doi          = {10.1007/978-3-031-98685-7\_14},
  bibsource    = {dblp computer science bibliography, https://dblp.org}
}


%% file: bib/smt-chc-sygus-co.bib
@article{UnnoTGK23,
  author       = {Hiroshi Unno and Tachio Terauchi and Yu Gu and Eric Koskinen},
  title        = {Modular Primal-Dual Fixpoint Logic Solving for Temporal Verification},
  journal      = {Proc. {ACM} Program. Lang.},
  publisher    = {{ACM}},
  volume       = {7},
  number       = {{POPL}},
  pages        = {2111--2140},
  year         = {2023},
  doi          = {10.1145/3571265},
}


%% file: bib/synthesis-abstraction.bib
@inproceedings{ChoiFPS22,
  author       = {Wonhyuk Choi and
                  Bernd Finkbeiner and
                  Ruzica Piskac and
                  Mark Santolucito},
  editor       = {Ranjit Jhala and
                  Isil Dillig},
  title        = {Can reactive synthesis and syntax-guided synthesis be friends?},
  booktitle    = {{PLDI} '22: 43rd {ACM} {SIGPLAN} International Conference on Programming
                  Language Design and Implementation, 2022},
  pages        = {229--243},
  publisher    = {{ACM}},
  year         = {2022},
  doi          = {10.1145/3519939.3523429},
}

@inproceedings{MaderbacherB22,
  author       = {Benedikt Maderbacher and
                  Roderick Bloem},
  editor       = {Alberto Griggio and
                  Neha Rungta},
  title        = {Reactive Synthesis Modulo Theories using Abstraction Refinement},
  booktitle    = {22nd Formal Methods in Computer-Aided Design, {FMCAD} 2022},
  pages        = {315--324},
  publisher    = {{IEEE}},
  year         = {2022},
  doi          = {10.34727/2022/ISBN.978-3-85448-053-2_38},
}

@inproceedings{RodriguezS23,
  author       = {Andoni Rodr{\'{\i}}guez and C{\'{e}}sar S{\'{a}}nchez},
  editor       = {Constantin Enea and Akash Lal},
  title        = {Boolean Abstractions for Realizability Modulo Theories},
  booktitle    = {Computer Aided Verification - 35th International Conference, {CAV} 2023},
  series       = {LNCS},
  volume       = {13966},
  pages        = {305--328},
  publisher    = {Springer},
  year         = {2023},
  doi          = {10.1007/978-3-031-37709-9_15},
}

@inproceedings{RodriguezS24,
  author       = {Andoni Rodr{\'{\i}}guez and C{\'{e}}sar S{\'{a}}nchez},
  editor       = {Michael J. Wooldridge and Jennifer G. Dy and Sriraam Natarajan},
  title        = {Adaptive Reactive Synthesis for {LTL} and {LTLf} Modulo Theories},
  booktitle    = {Thirty-Eighth {AAAI} Conference on Artificial Intelligence, {AAAI}
                  2024, Thirty-Sixth Conference on Innovative Applications of Artificial
                  Intelligence, {IAAI} 2024, Fourteenth Symposium on Educational Advances
                  in Artificial Intelligence, {EAAI} 2024},
  pages        = {10679--10686},
  publisher    = {{AAAI} Press},
  year         = {2024},
  doi          = {10.1609/AAAI.V38I9.28939},
}

@inproceedings{RodriguezGS24,
  author       = {Andoni Rodr{\'{\i}}guez and Felipe Gorostiaga and C{\'{e}}sar S{\'{a}}nchez},
  editor       = {S. Akshay and Aina Niemetz and Sriram Sankaranarayanan},
  title        = {Predictable and Performant Reactive Synthesis Modulo Theories via Functional Synthesis},
  booktitle    = {Automated Technology for Verification and Analysis - 22nd International Symposium, {ATVA} 2024},
  series       = {LNCS},
  volume       = {15055},
  pages        = {28--50},
  publisher    = {Springer},
  year         = {2024},
  doi          = {10.1007/978-3-031-78750-8_2},
}

@inproceedings{RodriguezGS25,
  author       = {Andoni Rodr{\'{\i}}guez and
                  Felipe Gorostiaga and
                  C{\'{e}}sar S{\'{a}}nchez},
  editor       = {Ruzica Piskac and
                  Zvonimir Rakamaric},
  title        = {Counter Example Guided Reactive Synthesis for {LTL} Modulo Theories\({}^{\mbox{*}}\)},
  booktitle    = {Computer Aided Verification - 37th International Conference, {CAV}
                  2025, Zagreb, Croatia, July 23-25, 2025, Proceedings, Part {IV}},
  series       = {Lecture Notes in Computer Science},
  volume       = {15934},
  pages        = {224--248},
  publisher    = {Springer},
  year         = {2025},
  url          = {https://doi.org/10.1007/978-3-031-98685-7\_11},
  doi          = {10.1007/978-3-031-98685-7\_11}
 }

@inproceedings{AzzopardiSPS25,
  author       = {Shaun Azzopardi and
                  Luca Di Stefano and
                  Nir Piterman and
                  Gerardo Schneider},
  editor       = {Ruzica Piskac and
                  Zvonimir Rakamaric},
  title        = {Full {LTL} Synthesis over Infinite-State Arenas},
  booktitle    = {Computer Aided Verification - 37th International Conference, {CAV}
                  2025, Zagreb, Croatia, July 23-25, 2025, Proceedings, Part {IV}},
  series       = {Lecture Notes in Computer Science},
  volume       = {15934},
  pages        = {274--297},
  publisher    = {Springer},
  year         = {2025},
  url          = {https://doi.org/10.1007/978-3-031-98685-7\_13},
  doi          = {10.1007/978-3-031-98685-7\_13}
}


%% file: bib/synthesis-other.bib
@inproceedings{FaellaP23, 
  author      = {Faella, Marco and Parlato, Gennaro}, 
  title       = {Reachability Games modulo Theories with a Bounded Safety Player}, 
  year        = {2023}, 
  publisher   = {AAAI Press}, 
  doi         = {10.1609/aaai.v37i5.25779},
  booktitle   = {Proceedings of the Thirty-Seventh AAAI Conference on Artificial Intelligence
                 and Thirty-Fifth Conference on Innovative Applications of Artificial Intelligence
                 and Thirteenth Symposium on Educational Advances in Artificial Intelligence}
}


%% file: bib/synthesis-symbolic.bib
@inproceedings{NeiderT16,
  author       = {Daniel Neider and
                  Ufuk Topcu},
  editor       = {Marsha Chechik and
                  Jean{-}Fran{\c{c}}ois Raskin},
  title        = {An Automaton Learning Approach to Solving Safety Games over Infinite
                  Graphs},
  booktitle    = {Tools and Algorithms for the Construction and Analysis of Systems
                          - 22nd International Conference, {TACAS} 2016},
  series       = {LNCS},
  volume       = {9636},
  pages        = {204--221},
  publisher    = {Springer},
  year         = {2016},
  doi          = {10.1007/978-3-662-49674-9_12},
}

@article{FarzanK18,
  author    = {Azadeh Farzan and
               Zachary Kincaid},
  title     = {Strategy synthesis for linear arithmetic games},
  journal   = {Proc. {ACM} Program. Lang.},
  volume    = {2},
  number    = {{POPL}},
  pages     = {61:1--61:30},
  year      = {2018},
  doi       = {10.1145/3158149},
}

@inproceedings{SamuelDK21,
  author    = {Stanly Samuel and
               Deepak D'Souza and
               Raghavan Komondoor},
  editor    = {Diomidis Spinellis and
               Georgios Gousios and
               Marsha Chechik and
               Massimiliano Di Penta},
  title     = {GenSys: a scalable fixed-point engine for maximal controller synthesis
               over infinite state spaces},
  booktitle = {{ESEC/FSE} '21: 29th {ACM} Joint European Software Engineering Conference
               and Symposium on the Foundations of Software Engineering},
  pages     = {1585--1589},
  publisher = {{ACM}},
  year      = {2021},
  doi       = {10.1145/3468264.3473126},
}

@inproceedings{SamuelDK23,
  author       = {Stanly Samuel and
                  Deepak D'Souza and
                  Raghavan Komondoor},
  title        = {Symbolic Fixpoint Algorithms for Logical {LTL} Games},
  booktitle    = {38th {IEEE/ACM} International Conference on Automated Software Engineering,
                  {ASE} 2023},
  pages        = {698--709},
  publisher    = {{IEEE}},
  year         = {2023},
  doi          = {10.1109/ASE56229.2023.00212},
}

@inproceedings{MaderbacherWB24,
  author       = {Benedikt Maderbacher and
                  Felix Windisch and
                  Roderick Bloem},
  editor       = {Tiziana Margaria and
                  Bernhard Steffen},
  title        = {Synthesis from Infinite-State Generalized Reactivity(1) Specifications},
  booktitle    = {Leveraging Applications of Formal Methods, Verification and Validation.
                  Software Engineering Methodologies - 12th International Symposium,
                  ISoLA 2024},
  series       = {LNCS},
  volume       = {15222},
  pages        = {281--301},
  publisher    = {Springer},
  year         = {2024},
  doi          = {10.1007/978-3-031-75387-9_17},
}


%% file: bib/verification-control-monitoring.bib
@inproceedings{BardinFLP03,
  author       = {S{\'{e}}bastien Bardin and Alain Finkel and J{\'{e}}r{\^{o}}me Leroux and Laure Petrucci},
  editor       = {Warren A. Hunt Jr. and Fabio Somenzi},
  title        = {{FAST:} Fast Acceleration of Symbolic Transition Systems},
  booktitle    = {Computer Aided Verification, 15th International Conference, {CAV} 2003},
  series       = {LNCS},
  volume       = {2725},
  pages        = {118--121},
  publisher    = {Springer},
  year         = {2003},
  doi          = {10.1007/978-3-540-45069-6_12},
}

@inproceedings{BardinFLS05,
  author       = {S{\'{e}}bastien Bardin and Alain Finkel and J{\'{e}}r{\^{o}}me Leroux and Philippe Schnoebelen},
  editor       = {Doron A. Peled and Yih{-}Kuen Tsay},
  title        = {Flat Acceleration in Symbolic Model Checking},
  booktitle    = {Automated Technology for Verification and Analysis, Third International
                  Symposium, {ATVA} 2005},
  series       = {LNCS},
  volume       = {3707},
  pages        = {474--488},
  publisher    = {Springer},
  year         = {2005},
  doi          = {10.1007/11562948_35},
}

@article{KroeningSTTW13,
  author       = {Daniel Kroening and Natasha Sharygina and Stefano Tonetta and Aliaksei Tsitovich and Christoph M. Wintersteiger},
  title        = {Loop summarization using state and transition invariants},
  journal      = {Formal Methods Syst. Des.},
  volume       = {42},
  number       = {3},
  pages        = {221--261},
  year         = {2013},
  doi          = {10.1007/s10703-012-0176-y},
}

@article{KincaidBCR19,
  author       = {Zachary Kincaid and
                  Jason Breck and
                  John Cyphert and
                  Thomas W. Reps},
  title        = {Closed forms for numerical loops},
  journal      = {Proc. {ACM} Program. Lang.},
  volume       = {3},
  number       = {{POPL}},
  pages        = {55:1--55:29},
  year         = {2019},
  url          = {https://doi.org/10.1145/3290368},
  doi          = {10.1145/3290368},
  bibsource    = {dblp computer science bibliography, https://dblp.org}
}

@article{PimpalkhareK24,
  author       = {Nikhil Pimpalkhare and
                  Zachary Kincaid},
  title        = {Monotone Procedure Summarization via Vector Addition Systems and Inductive
                  Potentials},
  journal      = {Proc. {ACM} Program. Lang.},
  volume       = {8},
  number       = {{OOPSLA2}},
  pages        = {1873--1899},
  year         = {2024},
  url          = {https://doi.org/10.1145/3689777},
  doi          = {10.1145/3689777},
  bibsource    = {dblp computer science bibliography, https://dblp.org}
}

@inproceedings{SolankiCLR24,
  author       = {Mayank Solanki and
                  Prantik Chatterjee and
                  Akash Lal and
                  Subhajit Roy},
  editor       = {Bernd Finkbeiner and
                  Laura Kov{\'{a}}cs},
  title        = {Accelerated Bounded Model Checking Using Interpolation Based Summaries},
  booktitle    = {Tools and Algorithms for the Construction and Analysis of Systems
                  - 30th International Conference, {TACAS} 2024, Held as Part of the
                  European Joint Conferences on Theory and Practice of Software, {ETAPS}
                  2024, Luxembourg City, Luxembourg, April 6-11, 2024, Proceedings,
                  Part {II}},
  series       = {Lecture Notes in Computer Science},
  volume       = {14571},
  pages        = {155--174},
  publisher    = {Springer},
  year         = {2024},
  url          = {https://doi.org/10.1007/978-3-031-57249-4\_8},
  doi          = {10.1007/978-3-031-57249-4\_8}
}

@inproceedings{SeryFS11,
  author       = {Ondrej Sery and
                  Grigory Fedyukovich and
                  Natasha Sharygina},
  editor       = {Kerstin Eder and
                  Jo{\~{a}}o Louren{\c{c}}o and
                  Onn Shehory},
  title        = {Interpolation-Based Function Summaries in Bounded Model Checking},
  booktitle    = {Hardware and Software: Verification and Testing - 7th International
                  Haifa Verification Conference, {HVC} 2011, Haifa, Israel, December
                  6-8, 2011, Revised Selected Papers},
  series       = {Lecture Notes in Computer Science},
  volume       = {7261},
  pages        = {160--175},
  publisher    = {Springer},
  year         = {2011},
  url          = {https://doi.org/10.1007/978-3-642-34188-5\_15},
  doi          = {10.1007/978-3-642-34188-5\_15}
  }


%% file: bib/verification-termination.bib
@inproceedings{ZhuK24,
  author       = {Shaowei Zhu and
                  Zachary Kincaid},
  editor       = {Arie Gurfinkel and
                  Vijay Ganesh},
  title        = {Breaking the Mold: Nonlinear Ranking Function Synthesis Without Templates},
  booktitle    = {Computer Aided Verification - 36th International Conference, {CAV}
                  2024, Montreal, QC, Canada, July 24-27, 2024, Proceedings, Part {I}},
  series       = {Lecture Notes in Computer Science},
  volume       = {14681},
  pages        = {431--452},
  publisher    = {Springer},
  year         = {2024},
  url          = {https://doi.org/10.1007/978-3-031-65627-9\_21},
  doi          = {10.1007/978-3-031-65627-9\_21},
  bibsource    = {dblp computer science bibliography, https://dblp.org}
}

@inproceedings{Frohn20,
  author       = {Florian Frohn},
  editor       = {Armin Biere and
                  David Parker},
  title        = {A Calculus for Modular Loop Acceleration},
  booktitle    = {Tools and Algorithms for the Construction and Analysis of Systems
                  - 26th International Conference, {TACAS} 2020, Held as Part of the
                  European Joint Conferences on Theory and Practice of Software, {ETAPS}
                  2020, Dublin, Ireland, April 25-30, 2020, Proceedings, Part {I}},
  series       = {Lecture Notes in Computer Science},
  volume       = {12078},
  pages        = {58--76},
  publisher    = {Springer},
  year         = {2020},
  url          = {https://doi.org/10.1007/978-3-030-45190-5\_4},
  doi          = {10.1007/978-3-030-45190-5\_4},
  bibsource    = {dblp computer science bibliography, https://dblp.org}
}

@inproceedings{FrohnG19,
  author       = {Florian Frohn and
                  J{\"{u}}rgen Giesl},
  editor       = {Clark W. Barrett and
                  Jin Yang},
  title        = {Proving Non-Termination via Loop Acceleration},
  booktitle    = {2019 Formal Methods in Computer Aided Design, {FMCAD} 2019, San Jose,
                  CA, USA, October 22-25, 2019},
  pages        = {221--230},
  publisher    = {{IEEE}},
  year         = {2019},
  url          = {https://doi.org/10.23919/FMCAD.2019.8894271},
  doi          = {10.23919/FMCAD.2019.8894271},
  bibsource    = {dblp computer science bibliography, https://dblp.org}
}

@inproceedings{FedyukovichZG18,
  author       = {Grigory Fedyukovich and
                  Yueling Zhang and
                  Aarti Gupta},
  editor       = {Hana Chockler and
                  Georg Weissenbacher},
  title        = {Syntax-Guided Termination Analysis},
  booktitle    = {Computer Aided Verification - 30th International Conference, {CAV}
                  2018, Held as Part of the Federated Logic Conference, FloC 2018, Oxford,
                  UK, July 14-17, 2018, Proceedings, Part {I}},
  series       = {Lecture Notes in Computer Science},
  volume       = {10981},
  pages        = {124--143},
  publisher    = {Springer},
  year         = {2018},
  url          = {https://doi.org/10.1007/978-3-319-96145-3\_7},
  doi          = {10.1007/978-3-319-96145-3\_7},
  bibsource    = {dblp computer science bibliography, https://dblp.org}
}

@inproceedings{BorrallerasBLOR17,
  author       = {Cristina Borralleras and
                  Marc Brockschmidt and
                  Daniel Larraz and
                  Albert Oliveras and
                  Enric Rodr{\'{\i}}guez{-}Carbonell and
                  Albert Rubio},
  editor       = {Axel Legay and
                  Tiziana Margaria},
  title        = {Proving Termination Through Conditional Termination},
  booktitle    = {Tools and Algorithms for the Construction and Analysis of Systems
                  - 23rd International Conference, {TACAS} 2017, Held as Part of the
                  European Joint Conferences on Theory and Practice of Software, {ETAPS}
                  2017, Uppsala, Sweden, April 22-29, 2017, Proceedings, Part {I}},
  series       = {Lecture Notes in Computer Science},
  volume       = {10205},
  pages        = {99--117},
  year         = {2017},
  url          = {https://doi.org/10.1007/978-3-662-54577-5\_6},
  doi          = {10.1007/978-3-662-54577-5\_6},
  bibsource    = {dblp computer science bibliography, https://dblp.org}
}

@inproceedings{UrbanGK16,
  author       = {Caterina Urban and
                  Arie Gurfinkel and
                  Temesghen Kahsai},
  editor       = {Marsha Chechik and
                  Jean{-}Fran{\c{c}}ois Raskin},
  title        = {Synthesizing Ranking Functions from Bits and Pieces},
  booktitle    = {Tools and Algorithms for the Construction and Analysis of Systems
                  - 22nd International Conference, {TACAS} 2016, Held as Part of the
                  European Joint Conferences on Theory and Practice of Software, {ETAPS}
                  2016, Eindhoven, The Netherlands, April 2-8, 2016, Proceedings},
  series       = {Lecture Notes in Computer Science},
  volume       = {9636},
  pages        = {54--70},
  publisher    = {Springer},
  year         = {2016},
  url          = {https://doi.org/10.1007/978-3-662-49674-9\_4},
  doi          = {10.1007/978-3-662-49674-9\_4},
  bibsource    = {dblp computer science bibliography, https://dblp.org}
}

@article{LeikeH15,
  author       = {Jan Leike and
                  Matthias Heizmann},
  title        = {Ranking Templates for Linear Loops},
  journal      = {Log. Methods Comput. Sci.},
  volume       = {11},
  number       = {1},
  year         = {2015},
  url          = {https://doi.org/10.2168/LMCS-11(1:16)2015},
  doi          = {10.2168/LMCS-11(1:16)2015},
  bibsource    = {dblp computer science bibliography, https://dblp.org}
}

@article{BagnaraMPZ12,
  author       = {Roberto Bagnara and
                  Fred Mesnard and
                  Andrea Pescetti and
                  Enea Zaffanella},
  title        = {A new look at the automatic synthesis of linear ranking functions},
  journal      = {Inf. Comput.},
  volume       = {215},
  pages        = {47--67},
  year         = {2012},
  url          = {https://doi.org/10.1016/j.ic.2012.03.003},
  doi          = {10.1016/J.IC.2012.03.003},
  bibsource    = {dblp computer science bibliography, https://dblp.org}
}

@inproceedings{ColonS01,
  author       = {Michael Col{\'{o}}n and
                  Henny Sipma},
  editor       = {Tiziana Margaria and
                  Wang Yi},
  title        = {Synthesis of Linear Ranking Functions},
  booktitle    = {Tools and Algorithms for the Construction and Analysis of Systems,
                  7th International Conference, {TACAS} 2001 Held as Part of the Joint
                  European Conferences on Theory and Practice of Software, {ETAPS} 2001
                  Genova, Italy, April 2-6, 2001, Proceedings},
  series       = {Lecture Notes in Computer Science},
  volume       = {2031},
  pages        = {67--81},
  publisher    = {Springer},
  year         = {2001},
  url          = {https://doi.org/10.1007/3-540-45319-9\_6},
  doi          = {10.1007/3-540-45319-9\_6},
  bibsource    = {dblp computer science bibliography, https://dblp.org}
}
